\documentclass[preprint]{aastex}
\usepackage{subfigure}

\newcommand{\be}{\begin{equation}}
\newcommand{\ee}{\end{equation}}

\newcommand{\pd}{\partial}
\newcommand{\msun}{M_{\odot}}

\newcommand{\mh}{m_{\rm{H}}}
\newcommand{\ergs}{\rm{erg} \rm{s}$^{-1}\,$}
\newcommand{\dotMFUV}{\dot{M}_{\rm{FUV}}}
\newcommand{\dotMX}{\dot{M}_X}
\newcommand{\fact}{\Lambda}

\begin{document} 

\title{\bf VISCOUS EVOLUTION AND PHOTOEVAPORATION OF 
CIRCUMSTELLAR DISKS DUE TO EXTERNAL FUV RADIATION FIELDS}  

\author{Kassandra R. Anderson$^{1,2}$, Fred C. Adams$^{1,2}$, and
  Nuria Calvet$^{2}$}  

\affil{$^1$Physics Department, University of Michigan, Ann Arbor, MI 48109} 

\affil{$^2$Astronomy Department, University of Michigan, Ann Arbor, MI 48109} 

\begin{abstract}
This paper explores the effects of FUV radiation fields from external
stars on circumstellar disk evolution.  Disks residing in young
clusters can be exposed to extreme levels of FUV flux from nearby OB
stars, and observations show that disks in such environments are being
actively photoevaporated.  Typical FUV flux levels can be factors of
$\sim 10^{2} - 10^{4}$ higher than the interstellar value.  These
fields are effective in driving mass loss from circumstellar disks
because they act at large radial distance from the host star, i.e.,
where most of the disk mass is located, and where the gravitational
potential well is shallow.  We combine viscous evolution (an
$\alpha$-disk model) with an existing FUV photoevaporation model to
derive constraints on disk lifetimes, and to determine disk properties
as functions of time, including mass loss rates, disk masses, and
radii.  We also consider the effects of X-ray photoevaporation from
the host star using an existing model, and show that for disks around
solar-mass stars, externally-generated FUV fields are often the
dominant mechanism in depleting disk material. For sufficiently large
viscosities, FUV fields can efficiently photoevaporate disks over the
entire range of parameter space.  Disks with viscosity parameter
$\alpha = 10^{-3}$ are effectively dispersed within $1-3$ Myr; for
higher viscosities ($\alpha = 10^{-2}$) disks are dispersed within
$\sim 0.25 - 0.5$ Myr. Furthermore, disk radii are truncated to less
than $\sim100$ AU, which can possibly affect the formation of planets.
Our model predictions are consistent with the range of observed masses
and radii of proplyds in the Orion Nebula Cluster.
\end{abstract}


\section{Introduction} 
\label{sec:intro} 

Circumstellar disks provide the material out of which planets,
asteroids, comets, and other solar system objects are engendered.  
An understanding of the time evolution of such disks is essential 
to explain the origins of our own Solar System, and the hundreds of 
additional exoplanetary systems.  Observed solar systems harbor
planets with a wide variety of masses, compositions, and orbital
elements; models of disk formation and evolution must be able to
account for this dispersion in the properties of extrasolar planets.

Observations show that circumstellar disks are dispersed relatively
quickly, often within $3-5$ Myr and nearly always within 10 Myr
\citep{hernandez}.  This time constraint introduces the need to
incorporate efficient mass depletion into existing disk theories (in
addition to viscous accretion onto the host star). One standard
explanation for the short observed disk lifetimes is photoevaporation
by radiation fields, either by the host star, or for disks in
sufficiently populated regions, by massive OB stars residing in the
surrounding cluster (see \citealt{adams_araa} and \citealt{armitage}
for general discussions of typical environments of young star-disk
systems and descriptions of disk evolution and sources of mass loss). 
Regardless of the source of the radiation fields, the effects on the
surface density profile can be significant, thereby influencing (and
possibly hindering) subsequent planet formation.

Many young star-disk systems reside in relatively isolated areas
(e.g., Taurus and Chamaeleon), with low stellar densities and few or
no nearby massive stars; in these regions, radiation from external
stars is not high enough to affect the disk structure, and the primary
source of mass loss (due to photoevaporation) is the host star itself.
However, in populated regions that contain OB stars, external fields
can be important and often provide more radiation than the host stars
\citep{armitage2000,fatuzzo2008,holden}.  Circumstellar disks can thus
be roughly separated into two classes based on the environment in
which they reside: Those living in isolation, where the disk is
affected primarily by the host star, and those living in groups where
external stars can influence the disk properties.  A better physical
understanding of the latter case is especially important because stars
often form in groups rather than in isolation \citep{lada2003}.  An
immediate example of this type of environment is the Orion Nebula
Cluster (ONC), which hosts the ``proplyds'' --- distorted disks that
appear to be actively photoevaporated by the Trapezium O stars,
especially $\theta^1$ C \citep{odell1993, henney, bally2000}.  This
idea is further supported by the fact that the average disk mass in
the ONC appears to be lower than in Taurus, and is an order of
magnitude lower than the minimum mass solar nebula \citep{eisner};
thus far, no observed disks in Orion seem to have masses greater than
0.034 $\msun$ \citep{mann2009}.  Disks close to the cluster center
(projected distances less than 0.3 pc from $\theta^1$ C) are inferred
to have the lowest masses \citep{mann2010}.  The luminosity of
$\theta^1$ C is high enough so that the proplyds are exposed to FUV
flux levels of $G_0 \sim 10^3 - 10^4$ (note that $G_0 = 1$ corresponds
to the typical flux in the interstellar medium, $1.6 \times 10^{-3}$
\ergs cm$^{-2}$ [\citealt{habing1968}]).  In contrast, T Tauri disks
in isolated regions experience FUV flux from the central star only of
order $G_0 \sim 10^2$ at a distance $r = 100$ AU \citep{bergin2003}.

Many previous studies of disk photoevaporation have focused on
internally-generated radiation fields, i.e., from the host star.  
EUV fields from the central star were first considered by 
\cite{hollenbach}; \cite{clarke} subsequently combined this EUV
radiation with a standard $\alpha$-disk model to explore the time
evolution.  This work showed that EUV fields can affect the disk
structure by creating an inner ``hole'' at a radius of order 10 AU,
but usually only on long time scales -- roughly a factor of ten longer
than observed disk lifetimes of $\sim 3 - 10$ Myr.  This conclusion
led to refinements of this initial effort \citep{alexander2006}, as
well as the consideration of other types of radiation from the host
star as agents of mass loss, including FUV radiation and X-rays
\citep{gorti,gorti2,ercolano2009}.  This latter work showed that both
types of radiation fields can deplete material at low to moderate
radii, clear gaps in the surface density, and produce disk lifetimes
that are consistent with observations.  Subsequent authors have
considered the relative importance of these photoevaporation agents.
For example, \cite{owen10,owen11a,owen11b} showed that the effects of
X-rays can prevent EUV photons from heating the disk and driving a
flow, and argued that X-rays could be the most important agent in disk
dispersal.

Alongside this work on radiation fields from the central star,
additional studies have investigated the effects of external radiation
fields (due to nearby stars) on protoplanetary disk evolution
\citep[e.g.,][]{johnstone, richling2000, storzer, adams}. The latter
paper presents detailed numerical simulations --- and derives analytic
approximations --- for the expected mass loss rates, and shows that
they can be comparable to (or even higher) than mass loss rates due to
internally-generated radiation fields.  Subsequent papers have coupled
these photoevaporation models with viscous accretion, in order to
evaluate the prospects for planet formation \citep{mitchell2010}, and
to compare the model predictions with the disk radii of observed
proplyds in the Orion Nebula Cluster \citep{clarke2007}.

The main goal of this paper is to combine existing photoevaporation models
with time-dependent disk models in order to derive constraints on disk
masses, radii, and mass-loss rates for a wide range of parameters.
Here we focus on photoevaporation from external FUV sources and use
the results of a previous study \citep{adams} to determine the mass
loss rates for different external FUV fluxes $G_0$.  A secondary goal
is to combine these external FUV radiation fields with X-rays from the
host star, and to determine the parameters where one source of
photoevaporation is more important than the other. Finally, this work
shows that disks trace out well-defined evolutionary tracks in the
mass-radius plane, where these tracks depend on the external (and
internal) radiation fields and the viscosity parameter $\alpha$.
These mass-radius planes (analogous to H-R diagrams for stellar
evolution) can be useful for comparing theoretical models for disk
evolution with observational data.

This paper builds upon previous work in a number of ways: Most prior
studies have focused on internally-produced radiation fields coupled
with viscous disk evolution, whereas this work develops similar models
for externally-illuminated disks. Although external radiation fields
have been considered \citep{armitage2000,clarke2007}, the parameter
space has not yet been fully explored and newly available data for
disk masses in the Orion Nebula Cluster \citep{mann2010} allows these
models to be more rigorously tested. This work presents results from a
large ensemble of disk simulations, which provide disk masses, disk
radii, and mass-loss rates as functions of time; we then compare the
results to current data.  Note that these results will be useful for
future observational comparisons.  We show that relatively modest
external FUV radiation fields ($G_0$ = 300) can sometimes dominate the
disk evolution by greatly reducing the lifetime of the disk, as well
as by truncating the outer edge.  This present work also differs from
previous papers in that we combine our external FUV model with an
existing model for X-ray evaporation from the host star
\citep{owen11b}, and we delineate the portion of parameter space where
one radiation field is more important than the other.  For example,
disks exposed to FUV flux levels $G_0 = 3000$ are nearly always
dominated by external, rather than internal, radiation fields.

One can understand the importance of external radiation as follows:
While the radiation flux from the host star is strongest in the inner
disk (due to its $r^{-2}$ dependence), the flux arising from external
sources is essentially constant over the entire disk (since the
distance to ``nearby'' stars is much larger than the dimensions of the
disk itself).  The effects from external fields are thus most
prominent in the outer regions, where mass is less tightly bound to
the host star (and where the radiation from the central star is
weakest).  As a result, evaporation from these two sources leads to
evolution that is qualitatively different: The radiation fields from
the host star tend to destroy disks from the ``inside out,'' whereas
fields from external sources work from the ``outside in.''  Since most
of the mass resides in the outer regions of the disk, external
radiation fields have the potential to deplete the overall disk mass
on relatively short time scales.

This paper is organized as follows.  Section 2 outlines our
formulation of the problem, including the basic equations and
assumptions.  After a brief discussion of numerical techniques,
Section 3 presents the results of our simulations for various FUV
fluxes, disk viscosities, and X-ray luminosities. We then compare the
predictions of our model with observed disk masses and radii in the
Orion Nebula Cluster, and show that the model predictions are in
agreement with the observations. We conclude, in Section 4, with a
summary of our results, a discussion of their implications, and some
suggestions for future work.

\section{Disk Evolution Model}
\subsection{Viscous Evolution}

In the case of purely viscous evolution, the diffusion equation for
the surface density $\Sigma$ has the well-known form 
\be
\frac{\pd \Sigma}{\pd t} = \frac{3}{r} \, \frac{\pd }{\pd r} \, 
\left[ r^{1/2}\,\frac{\pd}{\pd r}\, 
\left( \nu r^{1/2} \Sigma \right) \right] \,,
\ee
where $r$ is the radius in cylindrical coordinates and $\nu$ is the
viscosity \citep{pringle}.  We follow the standard $\alpha$-model
\citep{shakura1973}, where the viscosity $\nu$ is given in terms of
the expression 
\be
\nu = \alpha \frac{a_s^2}{\Omega} = 
\alpha \frac{k_b T}{\mh} {\left( \frac{r^3}{G M_*} \right)}^{1/2}, 
\ee 
where $\alpha$ is a dimensionless parameter that determines the
magnitude of viscosity, $a_s$ is the sound speed, $\Omega$ is the
Keplerian frequency, $T$ is the midplane temperature, $M_*$ is
the host star mass, and the fundamental constants $k_b$, $G$, and
$\mh$ have the usual meaning.  In this paper, we work in terms of
dimensionless variables, where the radial coordinate is scaled by
1 AU and time is scaled by 1 yr; we thus define dimensionless
variables through the transformation 
\be 
r \to {r \over 1 \, {\rm AU}} \qquad {\rm and} \qquad 
t \to {t \over 1 \, {\rm yr}} \,.
\ee 
We adopt a midplane temperature profile of the form 
$T = T_m r^{-1/2}$, which is consistent with irradiated accretion
disks \citep{dalessio99}. After making these substitutions, the
diffusion equation for the surface density takes the form 
\be
\frac{\pd \Sigma}{\pd t} = \frac{3 \beta}{r} \frac{\pd }{\pd r} 
\left [ r^{1/2}  \frac{\pd}{\pd r} \left(r^{3/2} \Sigma\right) \right ]. 
\label{viscous}
\ee
Notice that an effective viscosity coefficient $\beta$ has been 
introduced, where 
\be
\beta \equiv 2 \pi \alpha \left( {k_b T_m \over \mh} \right) 
\left( {r_0 \over G M_*} \right) \approx  
6 \times 10^{-3} \alpha \, \left( \frac{M_*}{\msun} \right)^{-1}
\left(\frac{T_m}{100 \ \rm{K}} \right) \,,
\label{beta}
\ee
where $r_0$ = 1 AU. 

Throughout this paper, standard boundary conditions for the surface
density $\Sigma$ are assumed, where the torque vanishes at the inner
boundary $r_1$ (so that $\Sigma(r_1)$ = 0), and where the disk is
allowed to expand freely at the outer boundary.  We choose an initial
surface density profile of the form 
\be
\Sigma(r, 0) =  \Sigma_0\frac{\exp{[-r/r_d]}}{r} \\
\approx \frac{M_d}{2 \pi r_d (1 - e^{-1})} \frac{\exp{[-r/r_d]}}{r},
\label{sigma0}
\ee 
where $\Sigma_0$ provides the normalization for the surface density,
and where $M_d$ and $r_d$ are the initial disk mass and radius. The
approximation equality becomes exact in the limit $r_1 \to 0$. For the
case of purely viscous evolution, the surface density approaches this
form for any choice of the initial profile; we thus adopt this form to
reduce the time needed for the surface density to reach its asymptotic
form.

In the presence of significant radiation fields, additional mass flows
are introduced through photoevaporation, and the surface density
profile decreases more rapidly with time. As a general rule,
photoevaporation causes mass to become unbound near (and outside) a
critical radius $r_g$ where the sound speed $a_s$ is comparable to the
escape velocity \citep[e.g.,][]{hollenbach,adams}. The critical radius
thus has the form 
\be
r_g = \frac{G M_*}{a_s^2} \approx 100 \, {\rm AU} \, 
\left( {M_* \over 1 M_\odot} \right) \, 
\left( {T \over 1000 \, {\rm K}} \right)^{-1} \,. 
\label{r_g}
\ee
The effects of radiation fields are included by introducing a sink 
term into equation (\ref{viscous}), which yields
\be
\frac{\pd \Sigma}{\pd t} = \frac{3 \beta}{r} \frac{\pd }{\pd r} 
\left[ r^{1/2}  \frac{\pd}{\pd r} (r^{3/2} \Sigma) \right ] - 
\dot{\Sigma}(r) \,,
\label{sigma}
\ee
where the radial dependence of the sink term $\dot{\Sigma}$ depends
upon the type of radiation field(s) under consideration.

\subsection{FUV Radiation Fields Due to External Stars}

\begin{figure}[p]
\centering 
\includegraphics[scale = 0.5]{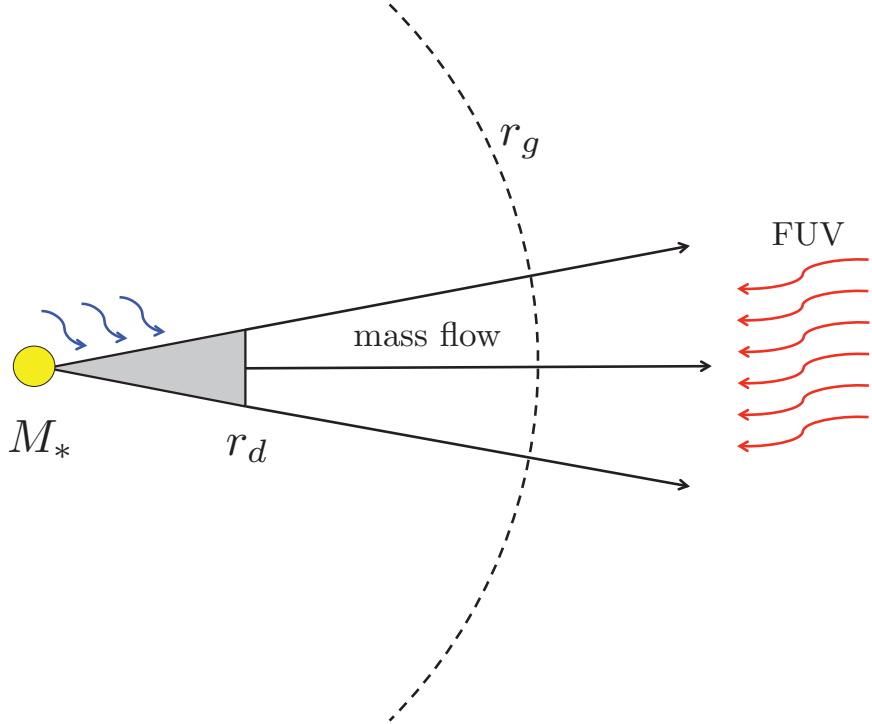}
\caption{Illustration of a subcritical disk (with $r_d < r_g$). The
disk (shown in grey) is viewed edge-on, with the gravitational radius 
$r_g$ included for reference (dashed arc).  Incident FUV radiation
from nearby OB stars heats the gas near the disk edge and causes a
nearly radial outflow of mass. Additional outflows in the vertical
direction (not shown) also contribute to the total photoevaporative
mass-loss, but are generally small compared to the radial flow. The 
host star provides additional radiation (shown in blue), which
includes FUV, EUV, and X-rays.  This paper focuses on the effects of
external FUV illumination, but also considers X-ray illumination from
the host star.}
\label{fig:schematic}
\end{figure}

We follow \cite{adams} to model FUV evaporation due to external stars,
who include detailed numerical calculations and semi-analytic
expressions for the total mass-loss rates as a function of the FUV
flux $G_0$.  They focus on ``subcritical disks,'' where the disk
radius $r_d$ is smaller than the gravitational escape radius $r_g$
(see equation [\ref{r_g}]). In this situation, FUV photons incident at
the outer disk edge penetrate radially inward and heat columns of gas.
A subsonic outflow of mass then develops at the disk edge and
accelerates outward; eventually the flow crosses a sonic radius $r_s$
(with $r_d < r_s < r_g$), where the flow becomes supersonic and mass
can freely escape.  Since the flow originates at large distances from
the host star, it can be approximated as spherically symmetric (with a
limited angular extent).  See Figure \ref{fig:schematic} for a
schematic illustration of this process.  Additional flow occurs in the
vertical direction, but the radial flow dominates the mass loss in the
subcritical ($r_d < r_g$) regime \citep{adams}. 

For the case of externally-illuminated disks, the critical radius
$r_g$ can be written in the form given by equation (\ref{r_g}). Note
that the given value of $r_g \sim 100$ AU differs from the typical
value quoted in the literature for internal evaporation from the host
star \citep[e.g.,][]{clarke}, where $r_g \sim 10$ AU is due to heating
by EUV photons.  This difference arises because the gas temperature in
regions of EUV ionization is quite high (and nearly isothermal), where
$T \sim 10^4$ K, and mass can become unbound at closer distances from
the host star.  In this case, however, the temperature $T$ in equation
(\ref{r_g}) refers to the temperature in the photodissociation region
(PDR) near the outer edge, where heating is due to FUV photons and the
temperature is lower ($T\sim10^2-10^3$ K).  This lower temperature is
reflected in a larger gravitational radius, where $r_g \sim 100$ AU.
Temperature profiles for the PDR are shown in Figure 2 of
\cite{adams}; these temperatures are a function of external FUV field
strength, as well as visual extinction $A_v$ and gas particle density.
The dependence of the FUV field strength is thus encapsulated into
$r_g$ implicitly through the temperature $T$.
 
For subcritical disks, with outer radius $r_d \lesssim r_g$, the
expected mass loss rates as a function of disk radius can be
approximated by 
\be
\dot{M}(r) = A r_g^{3/2} r^{1/2} \exp{[-r_g/2 r]},
\label{externalMdot}
\ee
where the parameter $A$ can be expressed as 
\be
A = C n_d a_s \langle \mu \rangle.
\ee
The constant $C$ is of order unity, $n_d$ is the gas particle density
at the disk edge, $a_s$ is the sound speed at the sonic radius, and
$\langle \mu \rangle$ is the mean particle mass (see Appendix A in
\citealt{adams} for more details).  Note that $\dot{M}(r)$ represents 
the total mass lost due to photoevaporation up to an outer radius, and
thus equation (\ref{externalMdot}) should be evaluated at $r = r_d$.
We can also write $\dot{M}$ as the integral 
\be
\dot{M} (r) = \int_{r_1}^r 2 \pi r' \dot{\Sigma} (r') dr'\,. 
\ee
By assuming conservation of mass, we can find the mass loss rate 
per unit area ($\dot{\Sigma}$) that is consistent with this total 
mass loss rate, i.e., 
\be
\frac{d\dot{M}}{d r} = 2 \pi r \dot{\Sigma} 
 = \frac{d}{dr} \left \{ A r_g^{3/2} r^{1/2} e^{-r_g/2 r}  \right\}.
\ee 
Expanding the derivative and solving for $\dot{\Sigma}$, 
we thus obtain the expression 
\be
\dot{\Sigma} (r) = \frac{A}{4 \pi} 
\left( \frac{r_g}{r} \right)^{3/2} 
\left[1 + \frac{r_g}{r} \right ] e^{-r_g/2 r}. 
\label{external}
\ee 
Equation (\ref{external}) captures the basic physics of external FUV
radiation and the radial dependence is dominated by the exponential
term.  The incident FUV flux is essentially constant (because the
distance to the source is much larger than the disk radius).  Gas
residing in the outer regions of the disk (where the gravitational
potential well is shallower) will thus experience the same FUV flux as
the gas in the inner disk, and can escape more easily.  Photons
incident at the disk edge can penetrate inward to some extent, heat
the gas, and initiate mass flow through the disk.  This flow, in
conjunction with mass conservation, gives rise to the radial
dependence of the sink term $\dot{\Sigma}$.  Since the incident
radiation is exponentially attenuated moving radially inward,
significant mass loss occurs primarily at larger radial distances, and
the disk shrinks from the outside in.
                              
To fully determine the mass-loss profile, the parameters $(r_g,A)$ in
equation (\ref{external}) must be specified.  Recall that the
dependence of the FUV field strength is implicitly included in the
expression for $r_g$ via the temperature $T$ in equation (\ref{r_g}).  
Here we fix these parameters by fitting a function of the form given
by equation (\ref{externalMdot}) to the numerical results shown in
Figure 4 of \cite{adams}, where the mass-loss rates $\dot{M}$ are
plotted as a function of outer disk radius and FUV radiation field.
The results of these (non-linear) curve fits yield an escape radius
$r_g = 357$ AU for a radiation field strength $G_0 = 300$, $r_g = 157$
AU for $G_0 = 3000$, and $r_g = 90$ AU for $G_0 = 30,000$.  Using
equation (\ref{r_g}), we can check that the results of these curve
fits correspond to reasonable temperatures by comparing with the PDR
temperature profiles in Figure 2 of \cite{adams}.  This comparison
shows that $T \approx 280, 640$, and 1110 K for $G_0 = 300, 3000$ and
$30,000$, which is consistent with the range of PDR temperatures
evaluated at visual extinctions of order unity, or, equivalently, to
evaluating the temperature near the base of the flow.  Note however
that for a given visual extinction, there is a range in the possible
temperatures, depending on the gas particle density.  This possible
range translates into uncertainties in $r_g$ and hence in the total
mass loss rates $\dot{M}$. More specifically, for a given FUV field
strength, $r_g$ can be higher or lower by a factor $\fact$, so that 
$r_g'= \fact r_g$ (where $0.5 \lesssim \fact \lesssim 2$). With this 
allowed variation, the possible change in our estimate for the total 
mass loss rate can be written in terms of the ratio 
\be
{\dot{M}' \over \dot{M}} \approx \fact^{3/2} e^{(1 - \fact)r_g/2r_d} 
\approx \fact^{3/2} e^{(1 - \fact)} \, , 
\ee
where we have assumed that the parameter $A$ does not change
appreciably with $r_g$; for the second approximate equality, we assume
that the typical disk radius is half the critical radius.  For the range
$0.5 \lesssim\fact\lesssim 2$, the bounds of $\dot{M}'/\dot{M}$ can be
estimated from graphical analysis, and we find that $0.5 \lesssim
\dot{M}'/\dot{M} \lesssim 1.1$.

\section{Results}

Equation (\ref{sigma}) can be numerically integrated upon
specification of the viscosity parameter $\beta$ (and thus $\alpha$,
see equation [\ref{beta}]), along with the strengths of the radiation
field(s) under consideration.  We employ an explicit finite-difference
integration scheme using a grid of 275 logarithmically-spaced points,
with an inner grid boundary at $r_1 \approx 0.05$ AU, and a large
outer boundary $r_{\rm{max}} \approx 50,000$ AU (the grid boundary is
thus much larger than the disks themselves).  To ensure accurate
results, we first ran simulations with all the sink terms set to zero
(i.e., purely viscous evolution) and checked that the numerical
results agreed with the known analytic solution \citep{lyndenbell}.
As an additional accuracy check, we also monitored the total angular
momentum of the disk, which is conserved provided that the total mass
lost at the innermost grid point is included.

We fix the host star mass at $M_* = 1 \msun$ and focus on an initial
disk mass $M_d = 0.1 M_*$.  This star-disk mass ratio is roughly the
largest that can occur without gravitational fragmentation of the disk
\citep{gammie2001,star90}.  Since one of the goals of this paper is to
calculate disk lifetimes in the presence of radiation fields, this
star-disk mass ratio will yield an upper limit on disk lifetimes.  We
choose $\alpha = 10^{-3}$ and $G_0 = 3000$ as the center of parameter
space, but explore wide ranges of both parameters, typically $10^{-4}
\leq \alpha \leq 10^{-2}$ and $300 \leq G_0 \leq 30,000$.  Recall that
in this formulation, the parameter $\beta$ in equation (\ref{beta})
determines the overall viscosity of the disk, which depends on the
disk midplane temperature $T_m$ at a distance $r = 1$ AU.  The disk
evolution is relatively insensitive to the choice of $T_m$, and we fix
this characteristic temperature at $T_m = 300$ K.  For the purposes of
this discussion, uncertainties in this characteristic temperature are
equivalent to uncertainties in $\alpha$. 

With the viscosity parameter $\alpha$, the radiation intensity $G_0$,
and the initial surface density profile specified, equation
(\ref{sigma}) can be integrated to solve for the surface density
profile of the disk as a function of time. We also consider other
quantities of interest, including the total disk mass, the outer disk
radius, the mass loss rates due to photoevaporation, and the mass
accretion rate onto the star.  Simulations are terminated after one of
the following criteria is met: [A] the disk has lost $99\%$ of its
original mass so that it has effectively been dispersed, or [B] more
than $10$ Myr have elapsed.

\subsection{Evaporation Due to FUV Radiation Fields From External Stars}

\begin{figure}[p]
\centering 
\includegraphics[scale = 0.75]{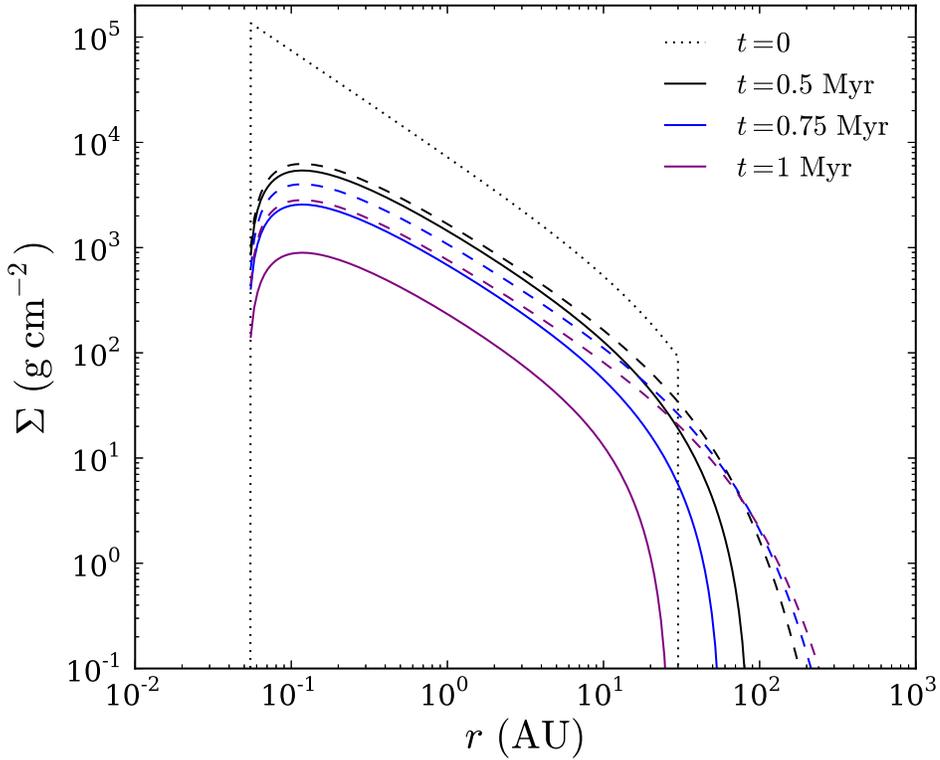}
\caption{Surface density profile for three distinct times, as labeled, 
with viscosity parameter $\alpha = 10^{-3}$.  The solid curves show
the disk immersed in an external FUV radiation field with strength
$G_0 = 3000$, and the dashed curves show purely viscous evolution (for
comparison). The initial surface density profile is also shown (given
by equation [\ref{sigma0}]).  For all the systems in this paper, the
initial disk mass and radius are $M_d = 0.1 \msun$ and $r_d = 30$ AU, 
respectively.  The initial surface density profile has discontinuities
at both the inner boundary ($r = 0.05$ AU) and at the initial outer
radius ($r = 30$ AU), but these discontinuities are quickly erased as
the disk diffuses.}
\label{fig:sigma}
\end{figure}

To begin, we focus on the effects of FUV radiation from external stars
(and postpone the discussion of internal radiation fields until the
next section).  To assess the effects of this radiation on the disk
evolution, we compare our results to a reference system with purely
viscous evolution (i.e., no additional radiation fields, so that the
sink terms in equation [\ref{sigma}] are set to zero).  The surface
density profile is shown in Figure \ref{fig:sigma}, where standard
values of the parameter space have been chosen to illustrate the basic
disk evolution ($\alpha=10^{-3}$ and $G_0=3000$).  The disk spreads on
a timescale that depends on radius and is determined by the viscosity
parameter.  As discussed in the previous section, the disk evolution
is affected most at large radii ($r \gtrsim r_g \sim 100$ AU).  The
main effect of this illumination by FUV photons is to truncate the
outer disk edge relative to the standard $\alpha$-disk solution. 
However, radiation affects the disk in other ways, especially by
decreasing the overall surface density after a significant fraction of
the initial disk mass has been lost.  Compared to the reference
system, which decreases exponentially at large radii, the surface
densities of disks exposed to FUV radiation decrease faster than an
exponential, and thus have more sharply-defined edges.

\begin{figure}[p]
\centering 
\includegraphics[scale = 0.75]{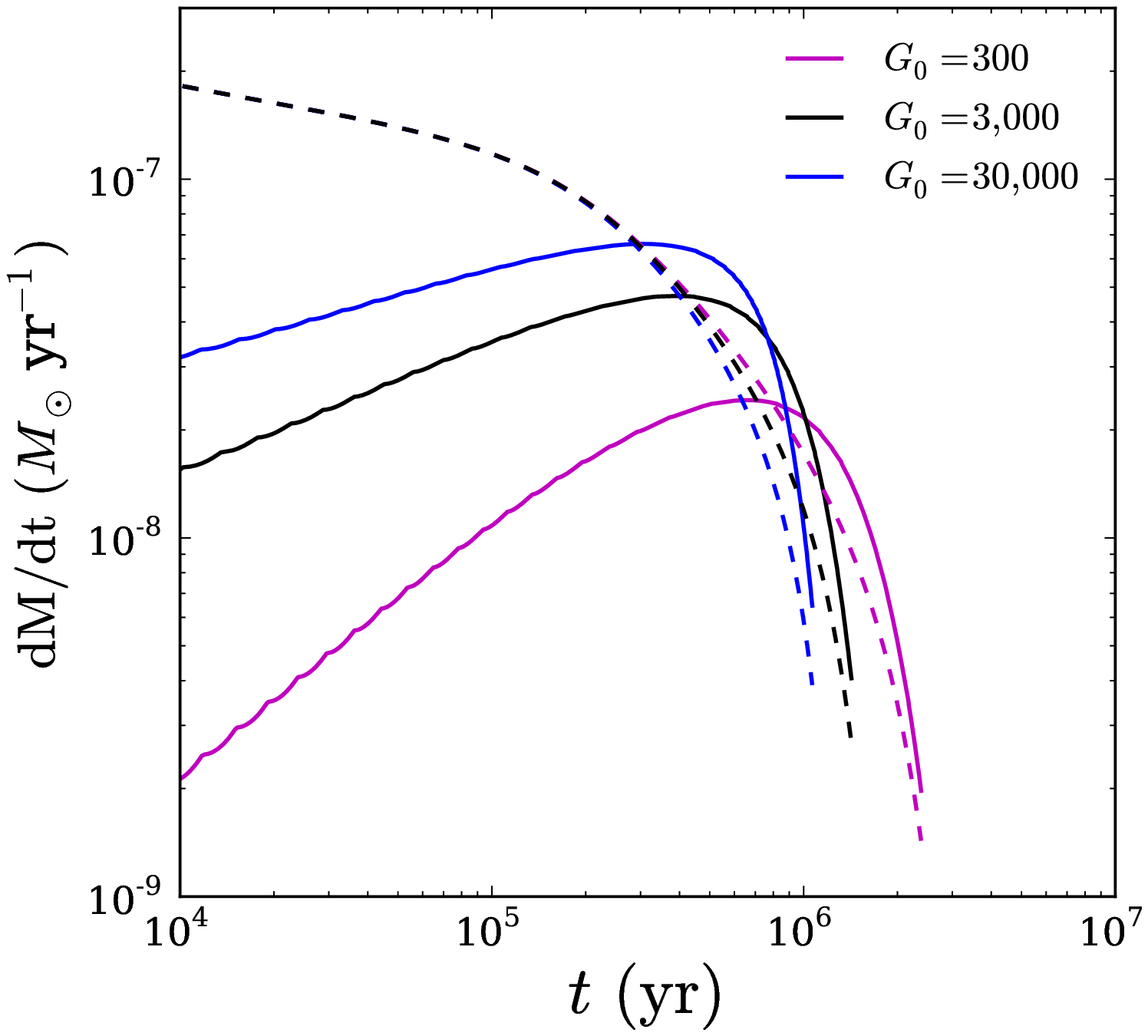}
\caption{Mass loss rate as a function of time due to photoevaporation
(solid curves) along with the mass accretion rate onto the star 
(dashed curves). These results were obtained for viscosity parameter 
$\alpha = 10^{-3}$.  For each value of the field strength $G_0$, the   
mass loss rate is initially relatively small, and steadily increases
as mass is transported outward towards regions of the disk where it
can escape.  The disk eventually reaches a steady-state, where the
time scale for mass loss is comparable to the viscous timescale. } 
\label{fig:mdot}
\end{figure}

Figure \ref{fig:mdot} shows the mass loss rate $\dot{M}$ through time
for disks with $\alpha = 10^{-3}$ and FUV fluxes $G_0=300,3000$, and
$30,000$.  The solid curves show the mass loss from photoevaporation
and the dashed curves show the usual mass accretion rate onto the host
star. At early times, the mass loss due to photoevaporation is
relatively low because the disk is compact (recall that at $t = 0$, we
have chosen $r_d = 30$ AU $ \ll r_g$).  As the disk diffuses outward,
the mass loss rate steadily increases, reaches a peak, and
then decreases again. Eventually, the time scale for mass loss due to
photoevaporation is comparable to the time scale for viscous evolution,
and the system reaches a steady state, where mass is steadily
evaporated as it is transported to the outer portions of the disk.  
As expected, the mass loss rates induced through photoevaporation
depend sensitively on the strength of external radiation field
strength; the mass accretion rate does not show this dependence,
however, because the inner portions of the disk are well-shielded from
the effects of the external radiation fields.

\begin{figure}[p]
\centering 
\includegraphics[scale = 0.75]{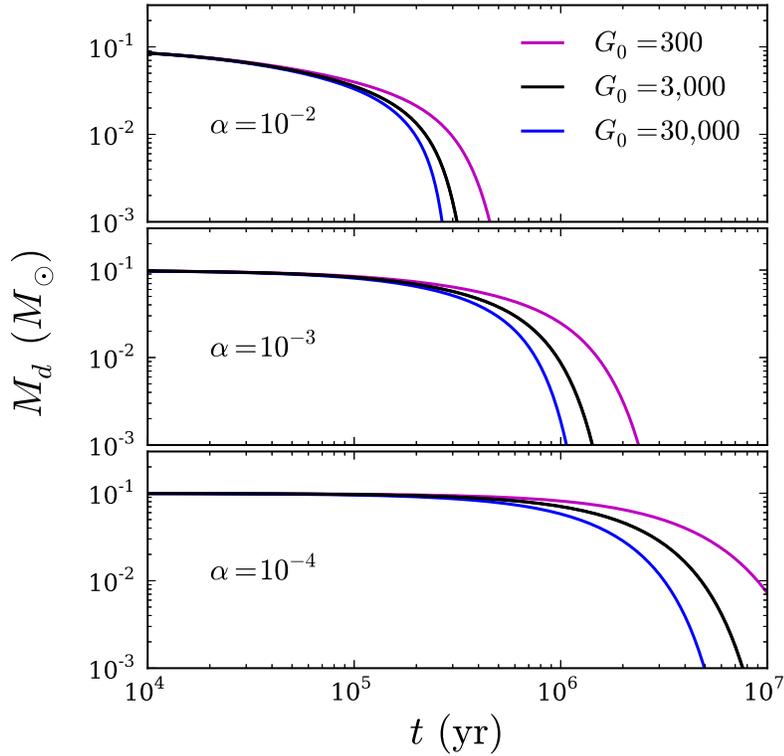}
\caption{Disk mass as a function of time for three values of $\alpha$ 
and $G_0$, as labeled.  Disks with high viscosity ($\alpha=10^{-2}$,
top panel) evolve quickly and are depleted within 0.5 Myr.  Disks with
moderate viscosity ($\alpha=10^{-3}$, middle panel) evolve more slowly
and survive somewhat longer, but are nonetheless dispersed in less
than $\sim 2.5$ Myr. Disks with low viscosity ($\alpha=10^{-4}$,
bottom panel) diffuse more gradually and can survive for at least 
$5$ Myr, and even longer than 10 Myr if the external field strength
is relatively low.}
\label{fig:mass}
\end{figure}

In comparison to the case of isolated systems, the lifetimes of
externally illuminated disks can be drastically shortened.  For the
particular disk shown in Figure \ref{fig:sigma}, most of the mass has
been lost after $\sim 1$ Myr, through both photoevaporation and
accretion onto the star.  At first glance, one might expect the disk
lifetime to be determined primarily by the radiation field strength
$G_0$. In practice, however, the disk lifetime is intimately linked to
the relative strengths of $G_0$ and the viscosity parameter $\alpha$.
For any radiation intensity, significant mass loss cannot occur unless
the disk expands enough so that mass reaches the outer (shallower)
portions of the gravitational potential well of the host star; when
this condition is met, the gas can freely escape. Thus, for efficient
mass loss, externally-illuminated disks must have sufficiently high
viscosity to cause enough mass to be transfered outward where it can
become unbound.  Figure \ref{fig:mass} shows the total disk mass as a
function of time for $\alpha = 10^{-4},10^{-3}$, and $10^{-2,}$, and
$G_0 = 300, 3000$, and $30,000$.  Disks with high viscosity ($\alpha =
10^{-2}$) are effectively dispersed in less than 0.5 Myr; disks with
low viscosity ($\alpha = 10^{-4}$) live for at least 5 Myr, and, for
low FUV flux ($G_0$ = 300), can survive for longer than 10 Myr. Note
that numerical estimates of $\alpha$ based on MRI simulations yield 
values in the range $\alpha \sim 10^{-3} - 10^{-2}$ 
\citep[e.g.,][]{brandenburg}.  These results imply that long-lived
disks should not be common in populated star forming regions.

\begin{figure}[p]
\centering 
\includegraphics[scale = 0.75]{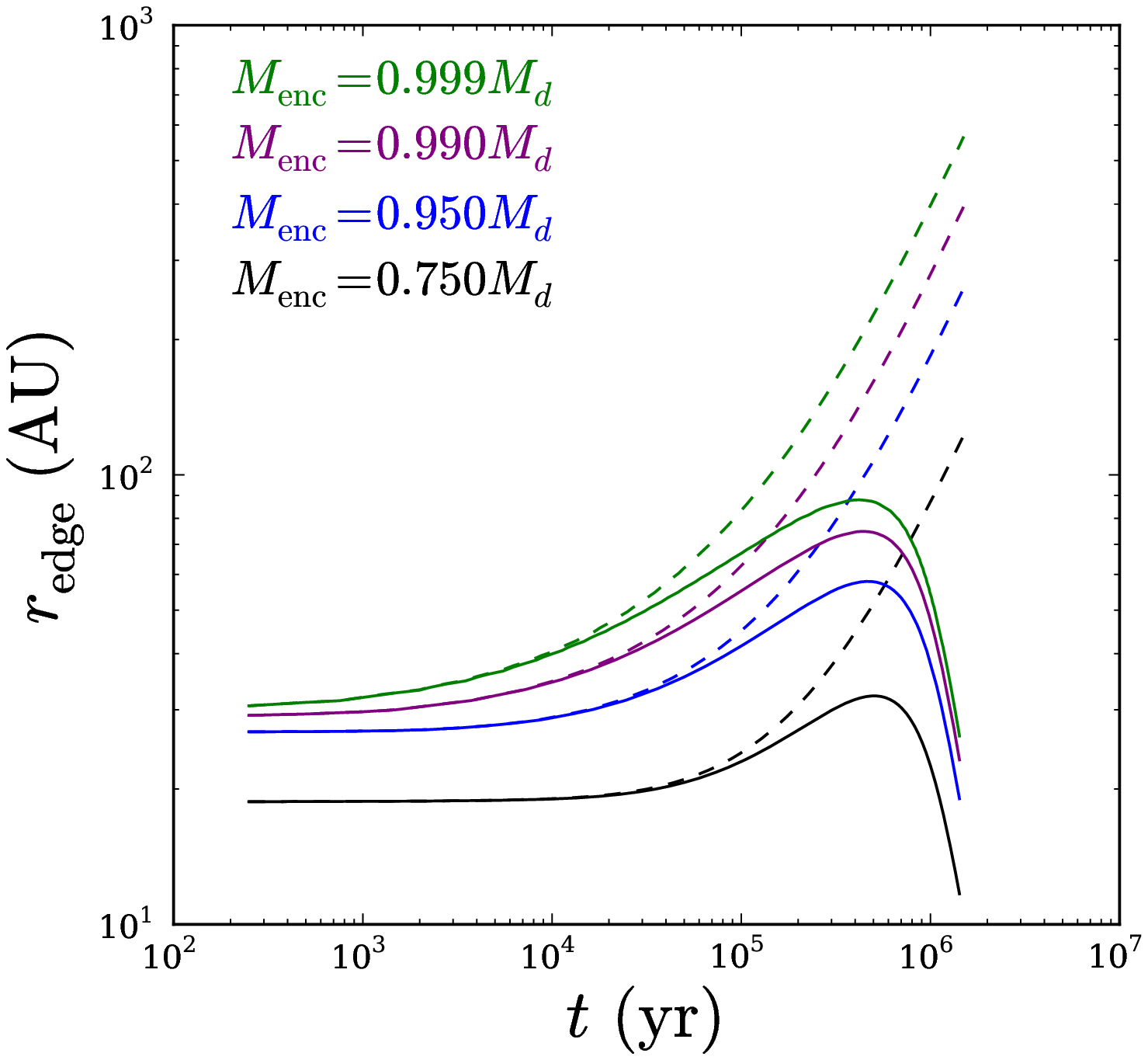}
\caption{Radius of our ``standard'' photoevaporating disk 
($G_0=3000$, $\alpha=10^{-3}$) as a function of time (solid curves), 
along with the results for a purely viscously evolving disk (dashed
curves). Each color corresponds to a different choice for the critical
fraction $f$ used to define the outer disk edge (see equation
[\ref{m_enc}]), as labeled.  The similar behavior of each curve
suggests that the choice of $f$ in defining the disk edge is not
particularly important; for the remainder of this paper, we adopt a
value $f = 0.99$.}
\label{fig:edge_f}
\end{figure}

\begin{figure}[p]
\centering 
\includegraphics[scale = 0.75]{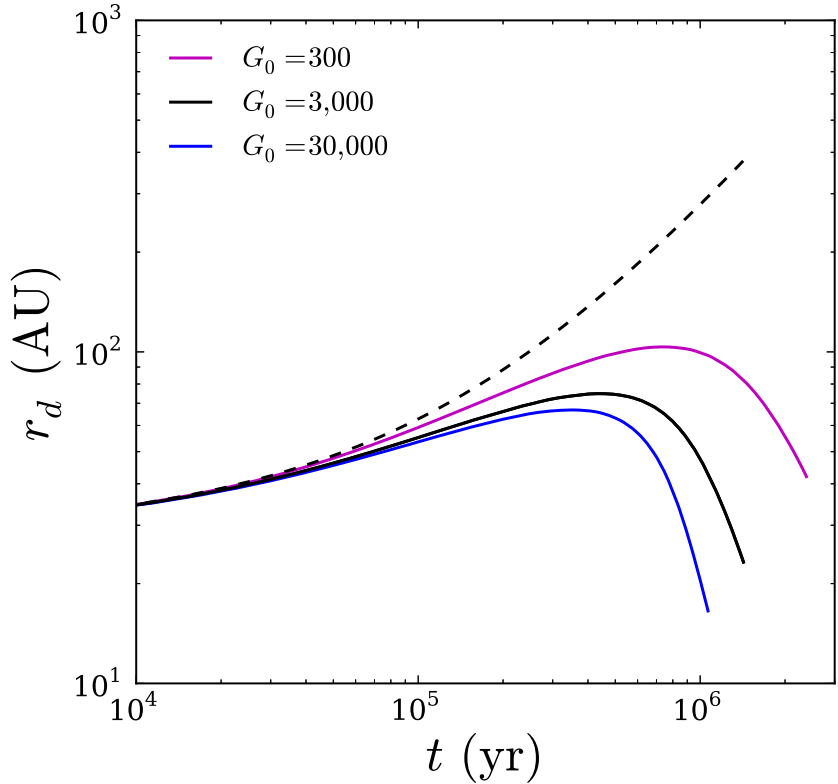}
\caption{Disk radius as a function of time for $\alpha = 10^{-3}$,  
and for radiation fields $G_0 = 300$ (top magenta solid curve), 
$G_0 = 3000$ (middle black solid curve), and $G_0 = 30,000$ (bottom
blue solid curve). The dashed curve, included for comparison, shows
the radius for a disk with no evaporation.  The contrast between the
power-law behavior of the dashed curve and the three solid curves
illustrates the truncation effects of the radiation fields.  The solid
curves (with evaporation) are initially smoothly increasing because
the viscous time scale is greater than the mass-loss time scale; mass
is thus transported outward faster than it can be evaporated.  As the
disk diffusion slows, mass is steadily evaporated as it travels
outward, and the disk radius begins to decrease.}
\label{fig:edge}
\end{figure}

\begin{figure}[p]
\centering 
\includegraphics[scale = 0.75]{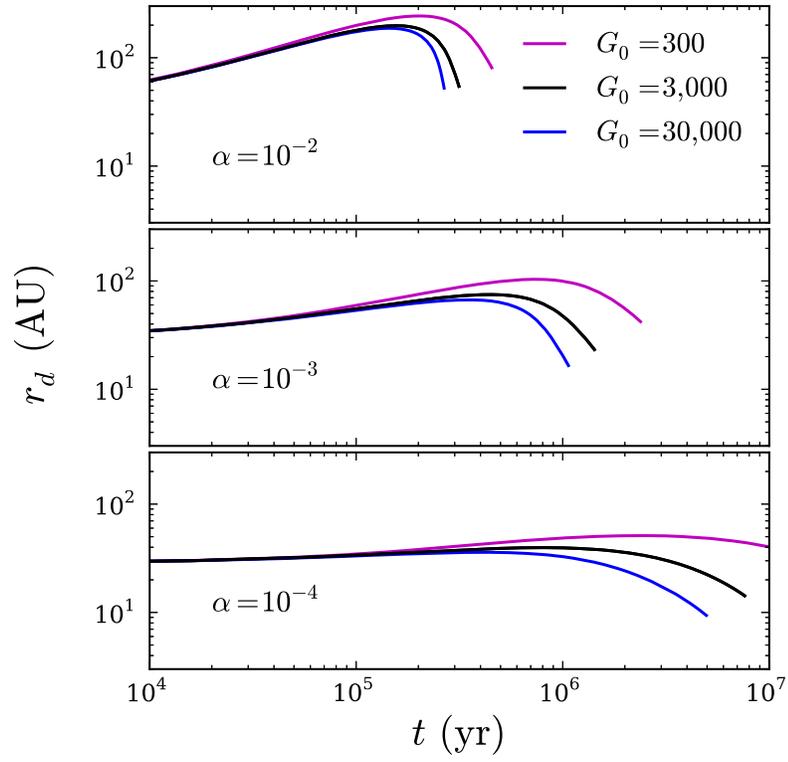}
\caption{Disk radius as a function of time (as in Figure \ref{fig:edge}) 
for three values of the viscosity parameter $\alpha$ and three values of 
the external radiation field $G_0$, as labeled.}
\label{fig:edge_all}
\end{figure}

In the absence of external radiation fields, the outer disk edge
increases with time as a power-law (and has no upper bound). Note that
the term ``outer disk edge'' itself is somewhat ambiguous, because the
standard $\alpha$-disk solution for the surface density decreases
exponentially at large radii, and thus does not have a clearly defined
outer boundary.  In order to get an estimate for the degree to which 
the disk is truncated, we define $r_d$ (where $r_d$ is a function of
time) as the radius where the enclosed mass is some critical fraction
of the total mass, i.e., 
\be
M_{\rm{enc}}(r_d,t) = \int_{r_1}^{r_d} 2 \pi r \Sigma dr = f M_d (t) =
f \int_{r_1}^{\infty} 2 \pi r \Sigma dr. 
\label{m_enc}
\ee 
Since the choice of the critical fraction $f$ is somewhat arbitrary,
we consider several values to ensure that the qualitative behavior of
$r_d(t)$ does not depend sensitively on the particular choice of $f$.
Figure \ref{fig:edge_f} shows the results for $f = 0.75, 0.95, 0.99$,
and $0.999$. Since the shape of the curves is similar for all four
cases, we can safely fix the critical fraction to be $f = 0.99$ for
the remainder of this paper.  This choice yields uncertainties in the
disk edge of order $10\%$ (where this uncertainy is due to the
definition of what we mean by the disk edge, and is not due to
inaccuracies in determining disk surface density profiles).  Figure
\ref{fig:edge} shows the disk edge $r_d$ as a function of time for
three values of $G_0$ and with fixed viscosity ($\alpha = 10^{-3}$).
Initially, the disk expands faster than it can be evaporated. As the
viscous time scale slows at later times, however, mass in the outer
disk becomes unbound as soon as it is transported outwards, and the
radius starts to decrease with time.  In all cases that include 
photoevaporation, after $\sim 0.5$ Myr have passed, the disk radii 
are noticably smaller than they would be in the absence of external
radiation. At longer times, $\sim1-2$ Myr, the disk radii become
smaller than their initial size. For this particular set of
parameters, the disk edge never increases beyond $r_d\approx100$ AU.
Figure \ref{fig:edge_all} shows the same results for different
viscosities.  The behavior is qualitatively similar for all
parameters.  Disks with the highest viscosities have the largest
maximum radii, but are dispersed on the shortest time scales.

\begin{figure}[p]
\centering 
\includegraphics[scale = 0.75]{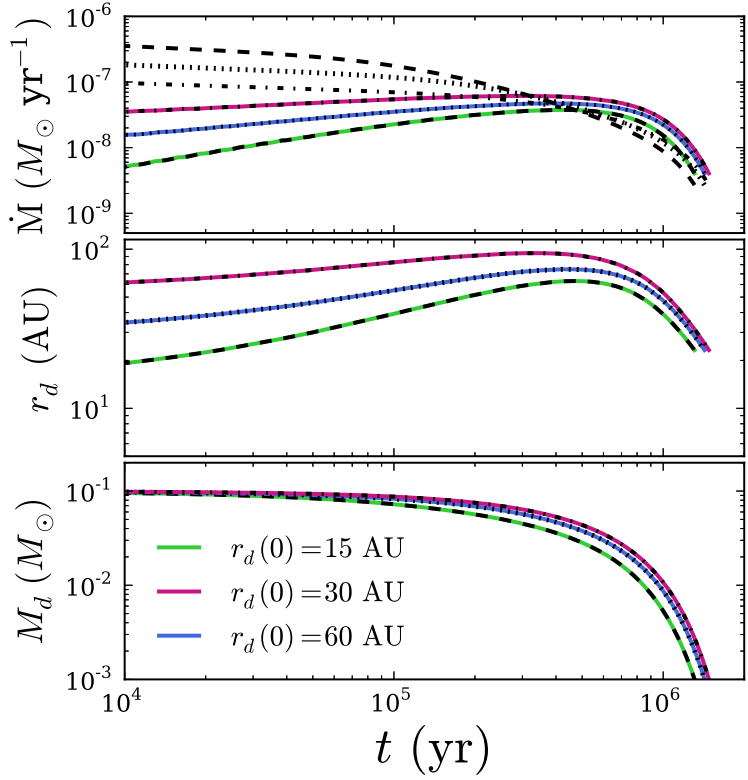}
\caption{Mass loss rate, disk radius, and disk mass through time for  
our standard parameters $G_0 = 3000$ and $\alpha = 10^{-3}$,
illustrating the effects of changing the initial radius $r_d(0)$.  The
colored curves correspond to the mass loss due to photoevaporation for
differing values of the starting radius, as labeled.  In the top
panel, the black dashed, dotted, and dash-dotted curves (from top to
bottom) show the mass accretion rate onto the host star for initial
radius $r_d(0)$ = 15, 30, and 60 AU respectively.  Note that although
the disk mass through time is almost independent of the initial
radius, disks with the smallest initial radii are dispersed on
slightly shorter time scales, because the mass accretion rate early 
in the disk evolution is the highest.} 
\label{fig:compare_radius}
\end{figure}

A potentially important parameter in these calculations is the initial
outer disk radius $r_d(0)$, because the extent of the disk determines
the mass loss due to evaporation (from equation [\ref{externalMdot}]).
Since our model focuses on the subcritical regime, the discussion is
restricted to initial radii less than the escape radius $r_g \sim 100$
AU.  We repeated our calculations using $r_d(0) = 15$ and 60 AU, i.e.,
within factors of two of our standard value $r_d(0) = 30$ AU.  The
resulting mass loss rates, radii, and masses are shown in Figure
\ref{fig:compare_radius}.  The disk mass as a function of time is
almost completely independent of the initial disk radius, and the mass
loss rate and disk edge converge to nearly the same value after
$\sim0.5$ Myr have elapsed. Note that disks with the smallest radii
are depleted the most quickly, in spite of the fact that the
photoevaporative mass loss is decreased (since less material is
located in the outer regions where it can escape).  This trend arises
because the mass accretion rate is initially highest for disks with
the smallest radii.  The main conclusion that can be drawn from Figure
\ref{fig:compare_radius} is that the initial disk radius is relatively
unimportant in the long-term evolution.

\begin{figure}[p]
\centering 
\includegraphics[scale = 0.75]{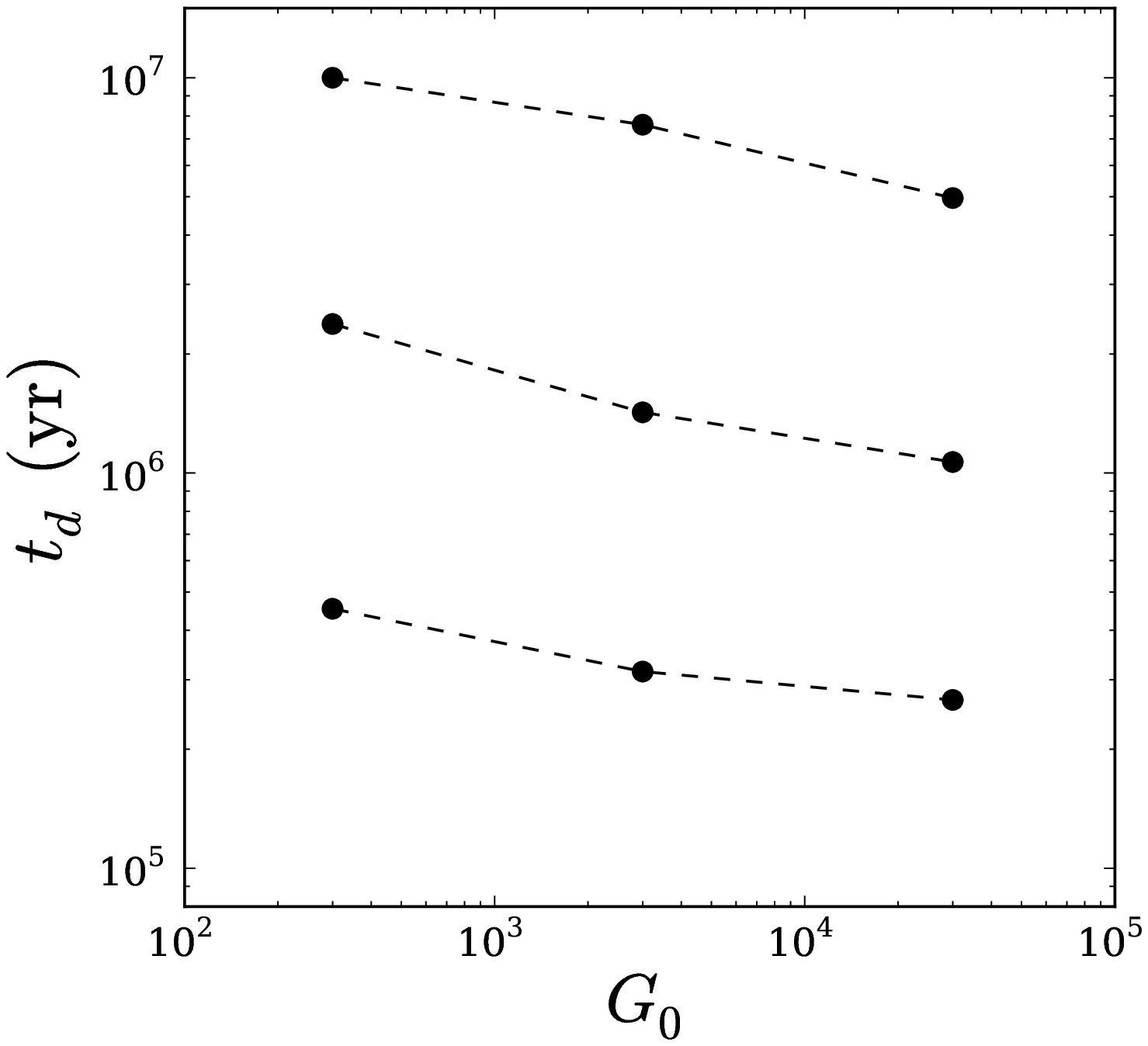}
\caption{Disk lifetime $t_d$ as a function of $G_0$ for 
$\alpha = 10^{-2}$, $10^{-3}$, and $10^{-4}$ (from bottom to top). 
We define $t_d$ to be the time needed for a disk to lose 99\% of its
original mass (if $t_d$ exceeds 10 Myr, however, we terminate the
simulation, and set $t_d = 10$ Myr to get a lower bound).  For a given
$G_0$, there is a range of disk lifetimes $t_d$, which depend on the
value of $\alpha$; this range spans over an order of magnitude, and 
shows the importance of not only the strength of the radiation field,
but also the amount of viscosity in the disk.  Notice that nearly all
disks are dispersed before 10 Myr have passed, and many in less than
2--3 Myr.}
\label{fig:G0_time}
\end{figure}

We can summarize our main results by plotting the disk lifetime $t_d$
as a function of the radiation field strength $G_0$ (for each value of
$\alpha$), as shown in Figure \ref{fig:G0_time}.  The Figure shows
that for fixed $G_0$, there exists a wide range of possible disk
lifetimes (spanning over an order of magnitude) depending on the
viscosity. This result highlights the importance of the viscosity on
disk dispersal.  Regardless of the radiation field strength, disks
with low viscosity ($\alpha \leq 10^{-4}$) can survive for relatively
long spans of time, essentially because mass is transported outward
slowly.  Nonetheless, notice that nearly all of the disks are
destroyed before 10 Myr have elapsed.  Furthermore, for expected
viscosities (with $\alpha \gtrsim 10^{-3}$), the disk mass is depleted
on rapid time scales $t_d \lesssim 2.5$ Myr.  These results show that
external FUV radiation fields can be important in disk dispersal, and
imply that disks with ages $t_d \gtrsim 3$ Myr should be rare or
nonexistent within richly populated clusters.

\subsection{Combined Effects of External FUV and Internal X-Ray Fields}

Given the results of the previous section, the next step is to include
photoevaporation due to radiation from the host star. In particular,
we want to identify the relative importance of internal and external
radiation fields, for varying choices of the other relevant parameters.  
\cite{owen11b} showed that out of the EUV, FUV, and X-ray radiation
fields generated by the host star, the X-rays are often the dominant
agent for disk dispersal.  As a result, we neglect FUV and EUV
radiation fields for this current assessment and consider only X-rays.
To model this X-ray photoevaporation, we use the numerical fits for
$\dot{\Sigma}$ provided by \cite{owen11b} (see Appendix B, equation
[B2]).  Figure \ref{fig:sdot} shows the radial dependence of the mass
loss profile, normalized to X-ray luminosities $L_X = 10^{29},
10^{30}, 10^{31}$ \ergs (dashed curves), along with the analytic
prediction for the mass loss profile for external FUV (equation
[\ref{external}]), normalized to $G_0 = 300, 3000, 30,000$ (solid
curves).

\begin{figure}[p]
\centering 
\includegraphics[scale = 0.75]{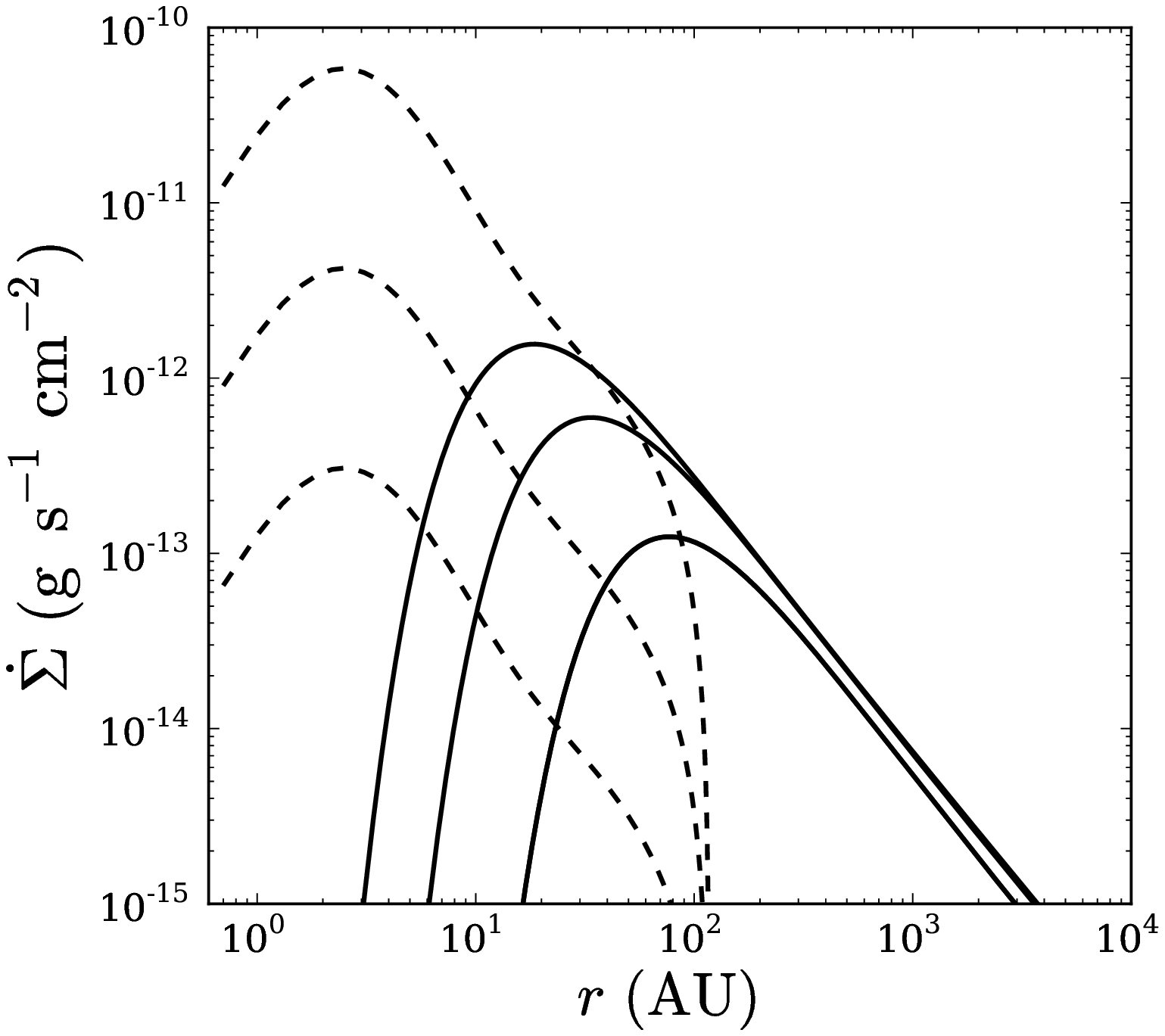}
\caption{Radial dependence of the sink terms $\dot{\Sigma}$.  The 
solid curves represent the analytic approximation for photoevaporation
from external stars (equation [\ref{external}]), taken from
\cite{adams}, normalized to flux levels $G_0 = 300, 3000$, and 30,000 
(bottom to top).  The dashed curves show the numerical fits for X-ray
evaporation from the host star, given by \cite{owen11b}.  The dashed
curves are normalized to X-ray luminosities $L_X = 10^{29}, 10^{30}$,
and $10^{31}$ erg s$^{-1}$ (bottom to top).}
\label{fig:sdot}
\end{figure}

\begin{figure}[p]
\centering 
\includegraphics[scale = 0.75]{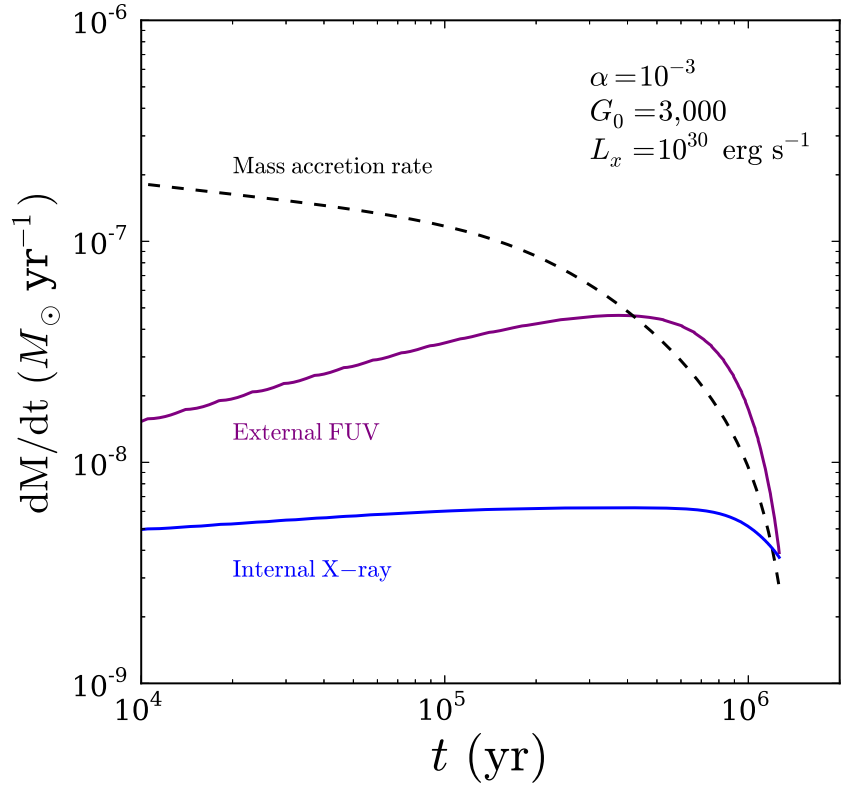}
\caption{Mass loss rates as functions of time, similar to those 
shown in Figure \ref{fig:mdot}, but with the inclusion of 
photoevaporation due to X-rays from the host star. The X-ray 
luminosity is $L_X = 10^{30}$ \ergs and the external FUV radiation 
field has $G_0$ = 3000. } 
\label{fig:mdot_Lx30}
\end{figure}

To start we fix the X-ray luminosity at $L_X = 10^{30}$ \ergs and look
at its effects on our standard disk with $\alpha = 10^{-3}$ and $G_0$
= 3000.  Figure \ref{fig:mdot_Lx30} shows the mass loss rates due to
each source, $\dotMFUV$ and $\dotMX$, along with the mass accretion
rate. Unlike the mass loss rate $\dotMFUV$ due to external FUV
radiation, the internal contribution $\dotMX$ does not increase
considerably with time as the disk spreads. The penetration of X-rays
from the host star is mostly confined to inner regions of the disk
(near the star), where a plentiful mass supply exists at the start of
the simulations.  Nonetheless, mass loss from X-rays increases slowly
as the disk spreads, so that a larger area of the disk can evaporate
(see the profiles in Figure \ref{fig:sdot}). At late times, the disk
radius shrinks, so that the mass loss from X-rays decreases.

\begin{figure}[p]
\centering 
\includegraphics[scale = 0.75]{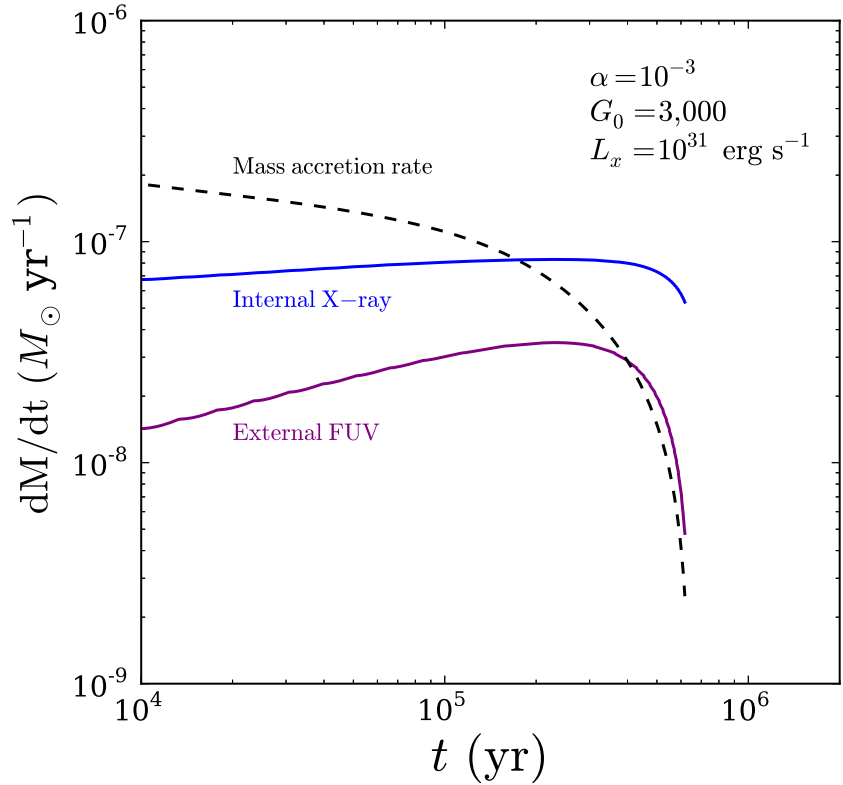}
\caption{Mass loss rates as functions of time, similar to those 
shown in Figure \ref{fig:mdot_Lx30}, but with $L_X = 10^{31}$ \ergs.
For this larger value of the X-ray luminosity, the mass loss rate due
to internal radiation $\dot{M}_x$ exceeds that due to external
radiation $\dotMFUV$ for all phases of disk evolution 
(compare with Figure \ref{fig:mdot_Lx30}). } 
\label{fig:mdot_Lx31}
\end{figure}

For the same disk and an elevated X-ray luminosity $L_X = 10^{31}$
\ergs (i.e., a factor of ten higher), the mass loss rates are shown in
Figure \ref{fig:mdot_Lx31}.  Note that this value for $L_X$ represents
the upper end of the possible range in X-ray luminosity for a solar
mass star \citep{flaccomio2012,guedel2007}, thereby giving an estimate
of the maximum possible effects of X-ray fields.  The Figure shows
that this increase in luminosity causes the mass loss from X-ray
evaporation to exceed that from external FUV fields; internal
radiation is thus important for this portion of parameter space.
Nonetheless, nearly all solar-type stars have X-ray luminosities
somewhat lower than $L_X$ = $10^{31}$ \ergs (e.g., see the X-ray
luminosity functions in \citealt{wang}).  As a result, for disks in
sufficiently populated clusters, we conclude that the dominant agent
for mass loss will generally be the external radiation fields.

\begin{figure}[p]
\centering 
\includegraphics[scale = 0.75]{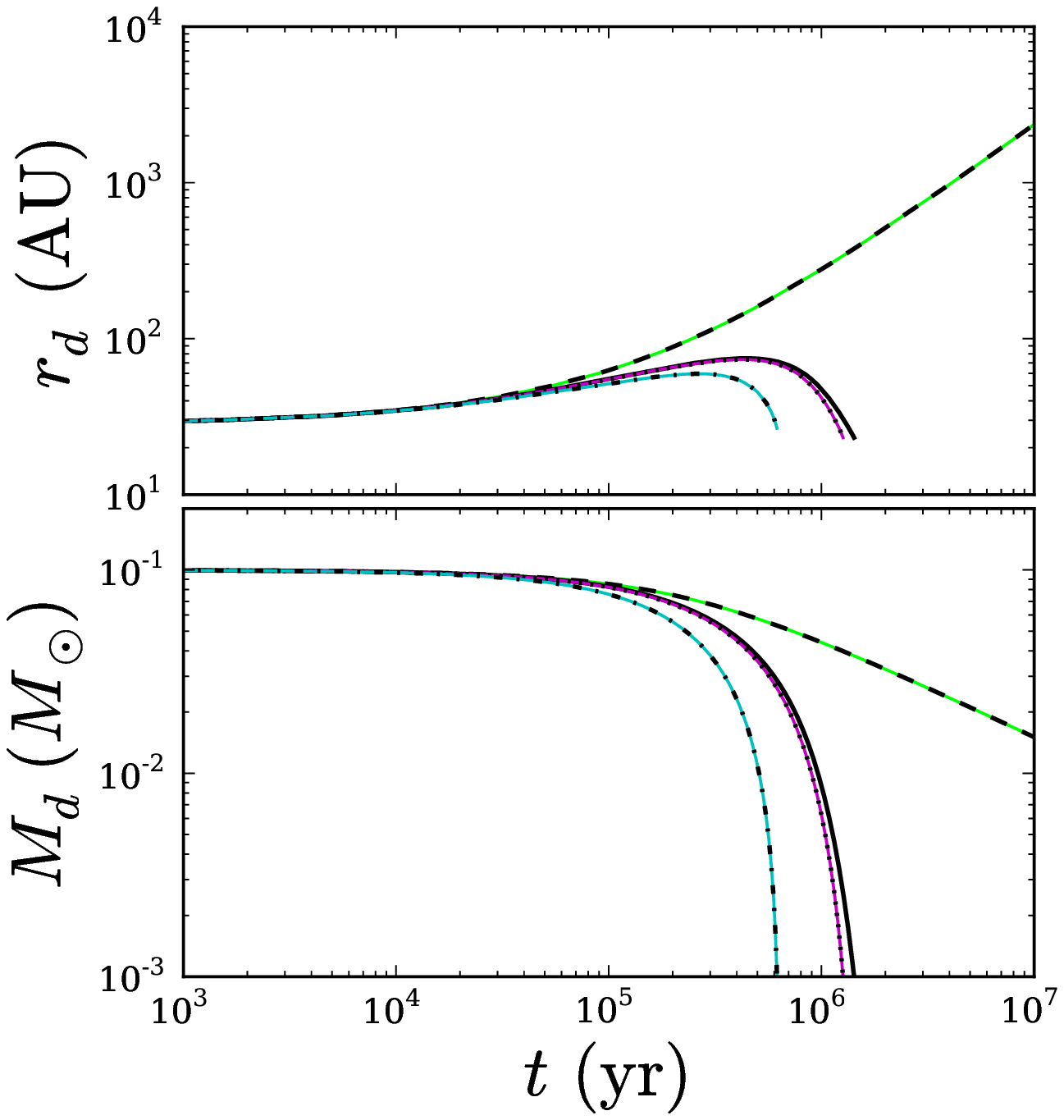}
\caption{Disk radius (top panel) and mass (bottom panel) as functions
of time for varying contributions from internal and external radiation
fields. All curves correspond to viscosity parameter $\alpha=10^{-3}$.
Green dashed curve shows results for a disk with no photoevaporation;
solid black curve corresponds to radiation fields $G_0$ = 3000 and
$L_X = 0$; magenta dotted curve corresponds to radiation fields $G_0$ =
3000 and $L_X = 10^{30}$ \ergs; and finally the cyan dash-dot curve
shows results for $G_0$ = 3000 and $L_X = 10^{31}$ \ergs. Notice that
the solid black and magenta dotted curves are nearly identical, i.e.,
an X-ray luminosity $L_X = 10^{30}$ \ergs does not significantly
influence disk evolution when coupled with an external FUV field.
However, increasing the luminosity to $L_X = 10^{31}$ \ergs shortens
the disk lifetime by nearly a factor of two.}
\label{fig:mass_rad_xray}
\end{figure}

Figure \ref{fig:mass_rad_xray} presents the time evolution for the
disk mass and disk radius for the varying strengths of the internal
radiation fields (determined by $L_X$) and external radiation fields
(determined by $G_0$). The effect of adding an X-ray contribution from
the host star with luminosity $L_X = 10^{30}$ \ergs to an external FUV
field is negligible, which is consistent with Figures
\ref{fig:mdot_Lx30} and \ref{fig:mdot_Lx31}.  Evaporation from host
stars with X-ray luminosities $L_X \leq 10^{30}$ \ergs is therefore
secondary in importance to evaporation by external FUV fields.
Increasing the X-ray luminosity to $L_X = 10^{31}$ \ergs, which
represents the upper end of the observed range, shortens the disk
lifetime by nearly a factor of two.

The discussion thus far has ignored the spectrum of the X-rays
irradiating the disk.  For completeness we note that as the X-ray
luminosity increases over its expected range ($L_X = 10^{29} -
10^{31}$ erg/s), the spectra are observed to become harder, and can
have a strongly penetrating ultrahard component during some flares
\citep{preibisch,getman}. 

\subsection{Comparison with Observed Proplyds in the Orion Nebula Cluster}

Next we test the previous developments by comparing the model
predictions with observed disks in the Orion Nebula Cluster (ONC).
These objects (often called the proplyds) are illuminated by radiation
from the four massive Trapezium stars, most notably $\theta^1$ Ori C,
a 40 $\msun$ O star near the cluster center.  Disks residing within
$\sim 2$ pc of the center are sufficiently illuminated so that the
results of this paper are applicable (but the observed proplyds have
much smaller projected distances).  Since the cluster radius is
estimated to be $r_c \approx 2.5$ pc \citep{hillenbrand}, the entire
population of disks in the cluster is potentially exposed to strong
radiation fields.

In this section, for the sake of definiteness, we consider only
external irradiation from FUV sources and neglect the effects of
X-rays from the host star. Using results from our model, including
both viscous evolution and external evaporation due to FUV radiation,
we construct ``evolutionary tracks'' in the mass-radius plane. Any
given disk must have a mass $M_d$ and $r_d$ at any given time, and our
model predicts the locus of points traced out in the ($r_d,M_d$) plane
as the system evolves. We can then compare these tracks with
observational data.  \cite{vicente2005} present diameters for 144
disks in the ONC.  Some diameters were measured by observing the dark
silhouette of the disk against the bright background (denoted
hereafter as the ``silhouette disks''), but most were inferred by
measuring the light from their own ionization fronts (IF) caused by
the hot Trapezium stars (hereafter the ``IF disks'').  The diameter of
the ionization front is somewhat larger than the true disk diameter;
\cite{vicente2005} attempted to correct for this effect, and estimated
that the IF diameter is approximately twice the disk diameter.  Their
estimated measurement errors of the IF diameters are $\sim 20$ AU;
however, the uncertainties in the disk diameters are somewhat larger,
because of uncertainties in the relationship between the ionization
front and the disk edge.  Note that their data set does not contain
any disks with radius $r_d < 30$ AU because such small disks were
below their resolution limit.  Disk masses for a subset of this sample
have been measured by \cite{mann2010} based on observations of their
sub-millimeter flux.  Calculating disk masses in this manner requires
assumptions about the dust opacity, the gas to dust ratio, and the
distance to the cluster \citep[see, for example,][]{williams_araa};
these assumptions lead to uncertainties in the disk masses.  Together,
the two data sets yield a sample of 28 disk systems in the ONC with
both measured masses and radii.  \cite{mann2010} also present upper
limits on disk masses for an additional 25 disks; including this
latter set of disks, the sample size increases to 53.

\begin{figure}[p]
 \centering
 \subfigure{
  \includegraphics[scale=0.48,trim = 20mm 6mm 25mm 15mm, clip]{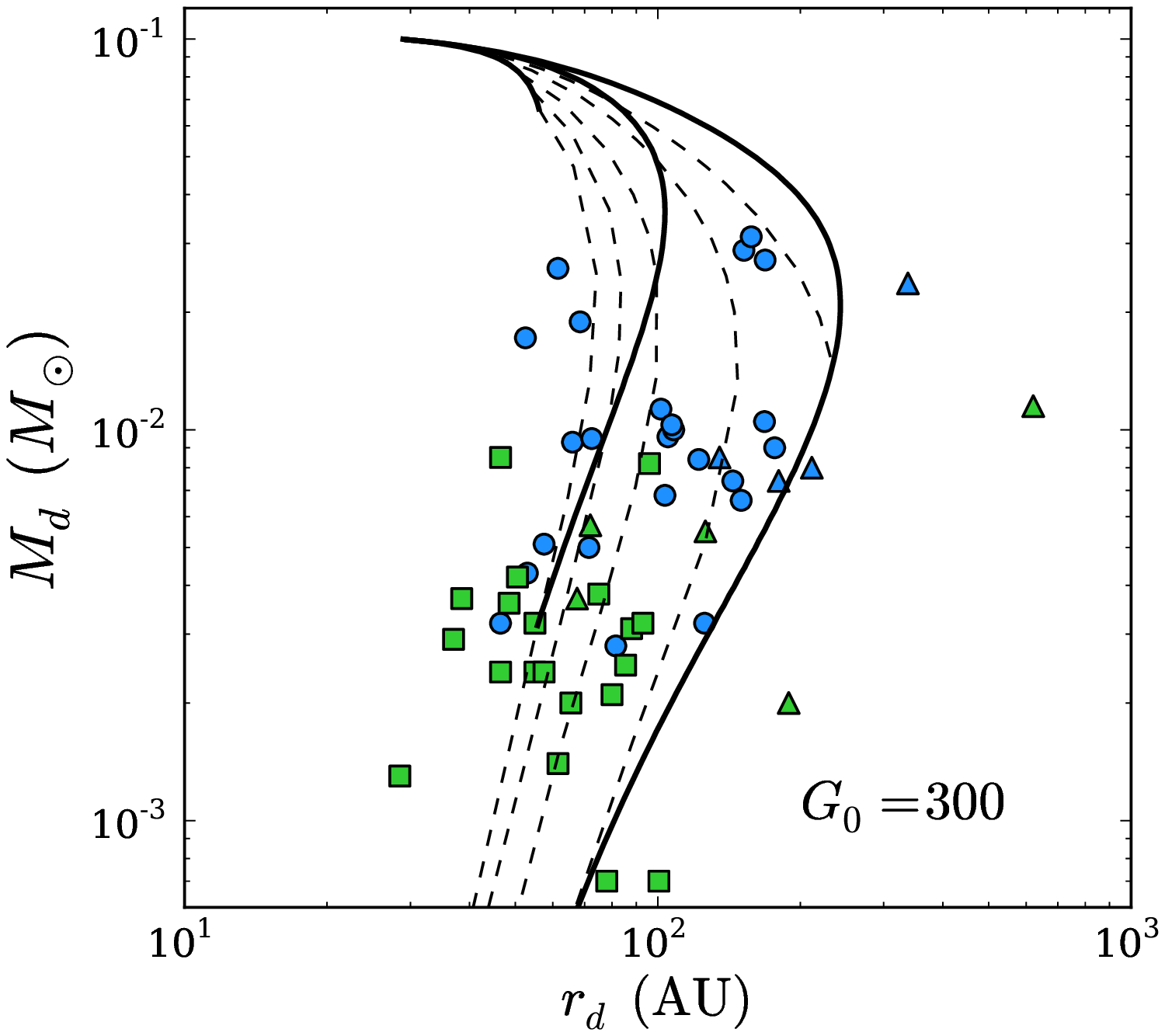}
 
   }
 \subfigure{
  \includegraphics[scale=0.48,trim = 20mm 6mm 25mm 15mm, clip]{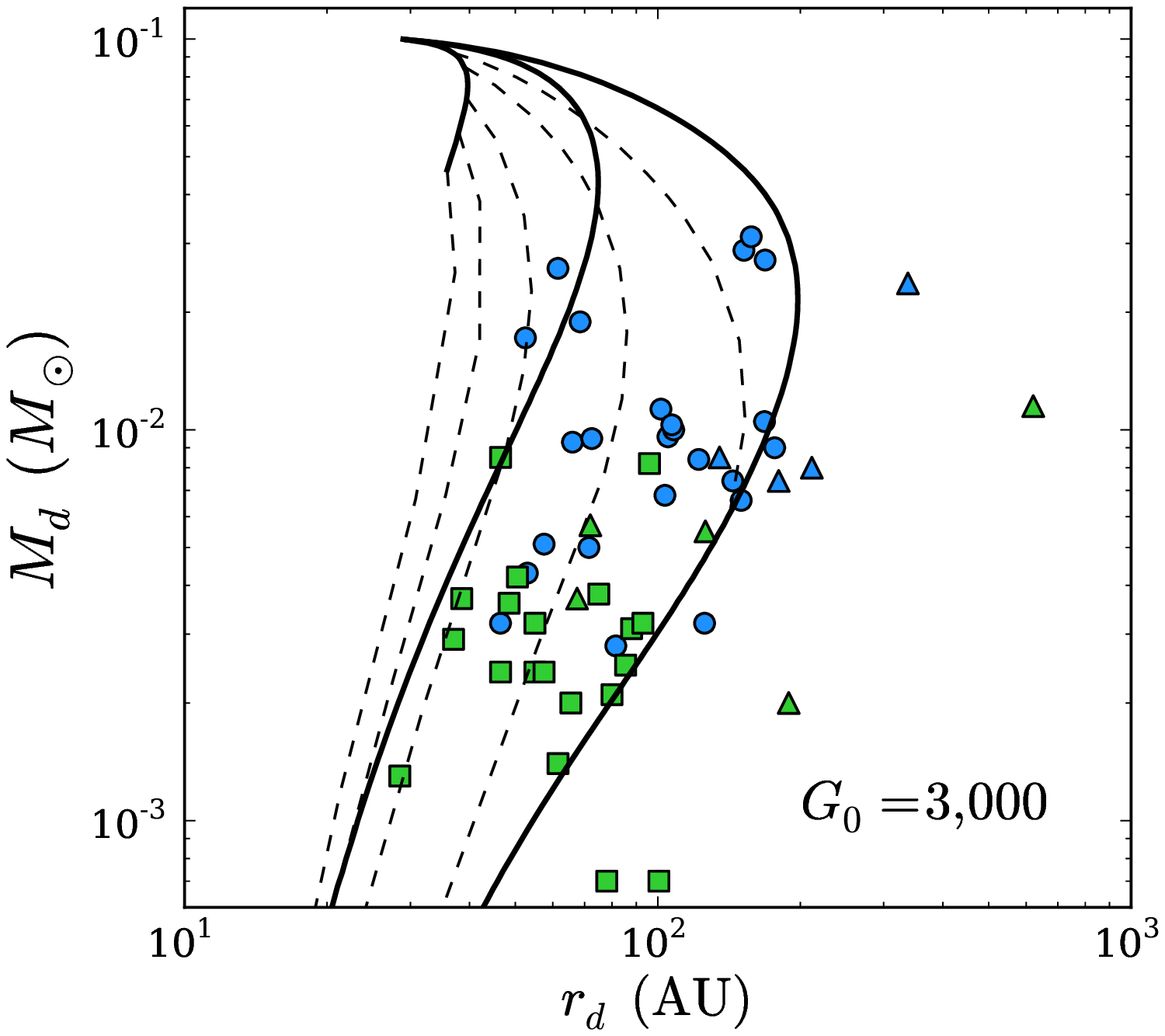}
 
   }
 \subfigure{
  \includegraphics[scale=0.48,trim = 20mm 6mm 25mm 15mm, clip]{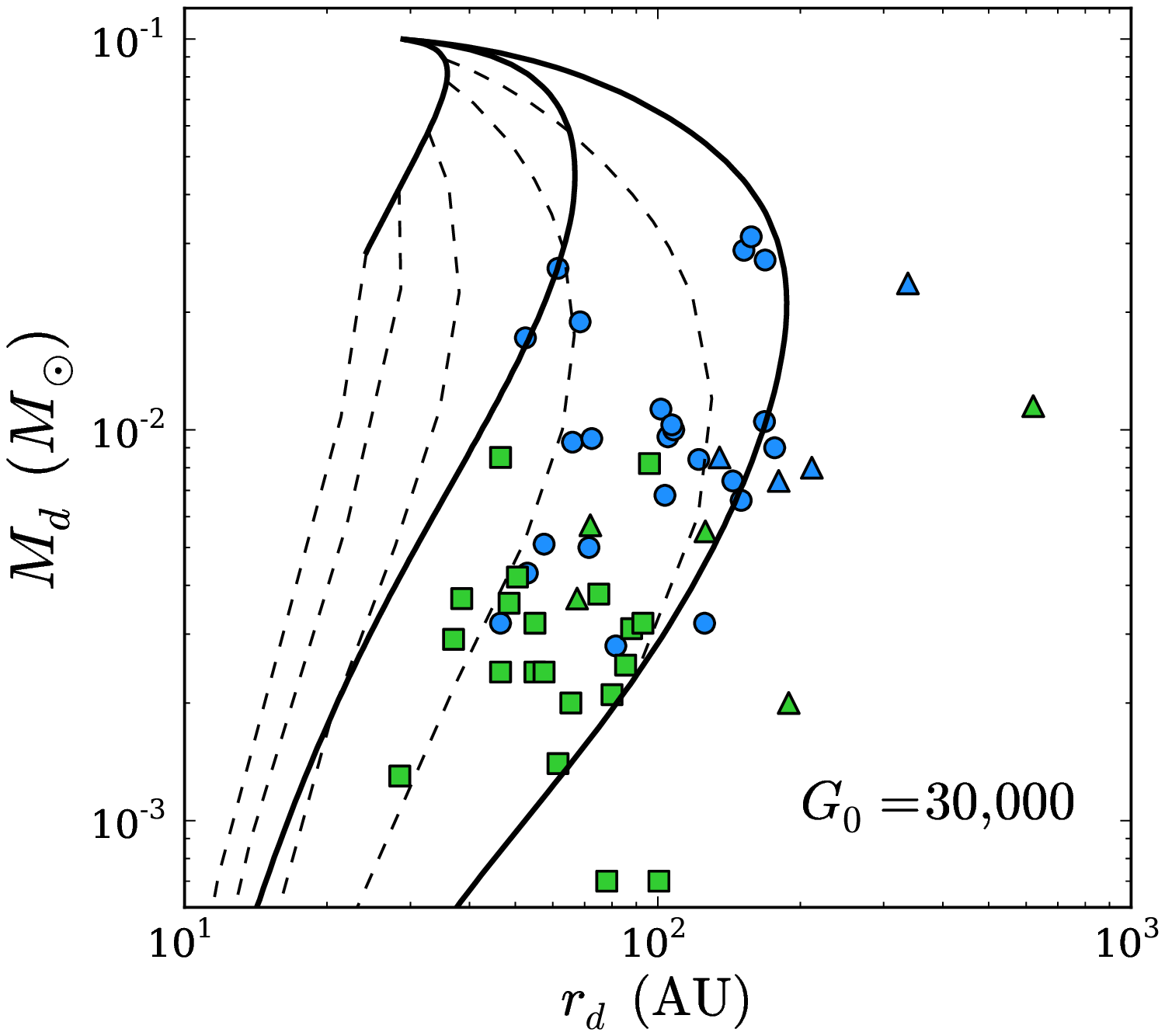}

   }

\caption{``Evolutionary tracks'' (solid curves) and ``isochrones''   
(dotted curves) in the plane of disk mass and radius $(r_d, M_d)$,   
along with observational data for disks in the Orion Nebula Cluster.  
The disk masses are taken from \cite{mann2010} and radii from
\cite{vicente2005}.  Blue points show disks with measured masses,
green points show data with mass upper limits. The triangles indicate
silhouette disks, whereas circles and squares indicate ionization
front disks (see text).  Each panel shows a different value of $G_0$,
as indicated. All of the evolutionary tracks start at the point $(r_d,
M_d)$ = $(30, 0.1)$ at $t = 0$ and travel downward in time.  These
evolutionary tracks are limited to times $t \leq 2$ Myr, to be consistent
with the inferred age of the ONC. In each panel, the three solid
curves show the disk evolution for different viscosities, with $\alpha
= 10^{-4}$ (left), $\alpha = 10^{-3}$ (middle), and $\alpha = 10^{-2}$
(right).  The dotted curves are isochrones, i.e., points at fixed
times, with times $t$ = 0.25, 0.5, 1.0, 1.5, and 2 Myr (from top 
to bottom). } 
\label{fig:isochrone}
\end{figure}

Figure \ref{fig:isochrone} compares the predictions of our model with
observed disk systems, where each panel shows a different value of the
external radiation field $G_0$ (from 300 to 30,000).  In each panel,
the bold solid curves show the disk evolution through time for
viscosity parameters $\alpha = 10^{-4}$ (left), $\alpha = 10^{-3}$
(middle), and $\alpha = 10^{-2}$ (right).  Note that the evolutionary
tracks are only plotted here for $t \leq 2$ Myr, because the observed
disks in the ONC are estimated to have ages of order 2 Myr or less.
As a result, the tracks for the lowest $\alpha$ values are truncated
(since disks with low viscosity evolve very slowly).  The dotted
curves connecting the solid curves are ``isochrones,'' i.e., curves
corresponding to constant values of time, but with varying sizes of
the viscosity parameter; here we have chosen viscosity parameters in
the range $10^{-4} \leq \alpha \leq 10^{-2}$ to represent the
approximate bounds of the parameter space.  The isochrones (shown as
dotted curves) mark evolutionary times of $t$ = 0.25, 0.5, 1.0, 1.5,
and 2 Myr (from top to bottom).  The blue data points show the 28
disks with estimated masses, and the green points show the 25 disks
with upper limits on their masses.  Triangles denote the silhouette
disks, while squares and circles indicate the IF disks.  As we argue
below, these models are consistent with disk ages that fall in the 
range $0.25 \lesssim t_d \lesssim 1.0$ Myr.

In general, the mass-radius data for observed disks is nicely bounded
by the ``edges'' of the expected parameter space. Essentially all of
the points in Figure \ref{fig:isochrone} are contained within the
region predicted for viscosity parameters in the range $\alpha$ =
$10^{-4}-10^{-2}$, radiation fields in the range $G_0$ = 300 --
30,000, and times $t < 2$ Myr. Furthermore, the best agreement between
the model and observations occurs for relatively intense radiation
fields with $G_0$ = 3000 or 30,000. This finding is sensible, given
that the disks reside in an environment (the ONC) containing many
massive stars.  Data points that fall outside of the model predictions
include many of the silhouette disks (marked by triangles).  This
result is not unexpected, because the silhouette disks are, on
average, farther from the cluster center, where the radiation fields
are weaker (and where larger disks can survive).  Note that there is a
degeneracy in the predicted disk mass and radius of our model; at a
given time, multiple combinations of $\alpha$ and $G_0$ can yield the
same mass and radius. As a result, we cannot unambiguously predict the
expected parameters for a particular observed disk. However, the range
of parameters for which the best agreement occurs are reasonable:
Figure \ref{fig:isochrone} indicates that most of the data points can
be understood using a limited range of parameter space, with a
preference for viscosity parameters $10^{-3}\leq\alpha\leq10^{-2}$,
and radiation fields with $G_0$ = 3000 or 30,000.  This range of
$\alpha$ values is predicted by many MRI simulations
\citep[e.g.,][]{brandenburg}, and this range of $G_0$ values is
typical for young clusters \citep{fatuzzo2008,holden}.  To move
forward, we need to break the degeneracy between the $\alpha$ and
$G_0$ values, which requires additional information; for example, it
would be useful to find the true distance of a disk from the cluster
center (rather than projected distance).

The results shown in Figure \ref{fig:isochrone} are based on a single
initial disk radius $r_d = 30$ AU. In practice, however, young disks
will display a range of radii, where the value depends on the initial
angular momentum and its evolutionary history.  To explore the effects
of varying initial disk sizes, we have repeated these calculations for
initial radii $r_d = 15$ and $r_d = 60$ AU, i.e., varying the disk
radius by a factor of two in both directions (see Section 3.1 and
Figure \ref{fig:compare_radius}). The resulting evolutionary plots
(not shown) are nearly identical to those in Figure
\ref{fig:isochrone}. This finding demonstrates that disk properties
(as a function of time) are relatively insensitive to the initial disk
radius, in agreement with previous work \citep{clarke2007}, as well as
the results shown in Figure \ref{fig:compare_radius}. 

Although the true distances of individual disks from the cluster
center in 3-dimensional space cannot be inferred with current data, we
can estimate the expected distribution of distances (and hence FUV
fluxes) for the entire sample.  Denoting the true distance as $r$ and
the projected distance as $d_p$, the (hidden) line-of-sight distance
$s$ is given by the expression 
\be
r^2 = d_p^2 + s^2 \, . 
\label{r}
\ee
Here, we allow the line-of-sight distance $s$ to be uniformly sampled 
within the range
\be
0 \leq s \leq r_c \sqrt{1 - d_p^2/r_c^2},
\label{s}
\ee
where $r_c$ is the cluster radius.  \cite{vicente2005} provide
projected angular distances for their sample of disks; we convert
these measurements to linear projected distances by assuming a
distance to the ONC of 414 pc \citep{menten2007}. We then set the
cluster radius $r_c = 2.5$ pc \citep{hillenbrand}, and repeatedly
sample $s$ for each observed disk in the range given by equation
(\ref{s}).  The dimensionless FUV flux then takes the form 
\be
G_0 = \frac{1} {F_0}\frac{L_{\rm{FUV}}}{4 \pi (d_p^2 + s^2)}, 
\ee 
with the typical interstellar flux level $F_0 = 1.6 \times 10^{-3}$
erg s$^{-1}$ cm$^{-2}$ \citep{habing1968}.  A cluster like the ONC
consisting of $N \sim 2000$ stars is expected to have an FUV
luminosity $L_{\rm{FUV}} \approx 2.46 \times 10^{39}$ \ergs
\citep{fatuzzo2008}.

\begin{figure}[p]
\centering 
\includegraphics[scale = 0.75]{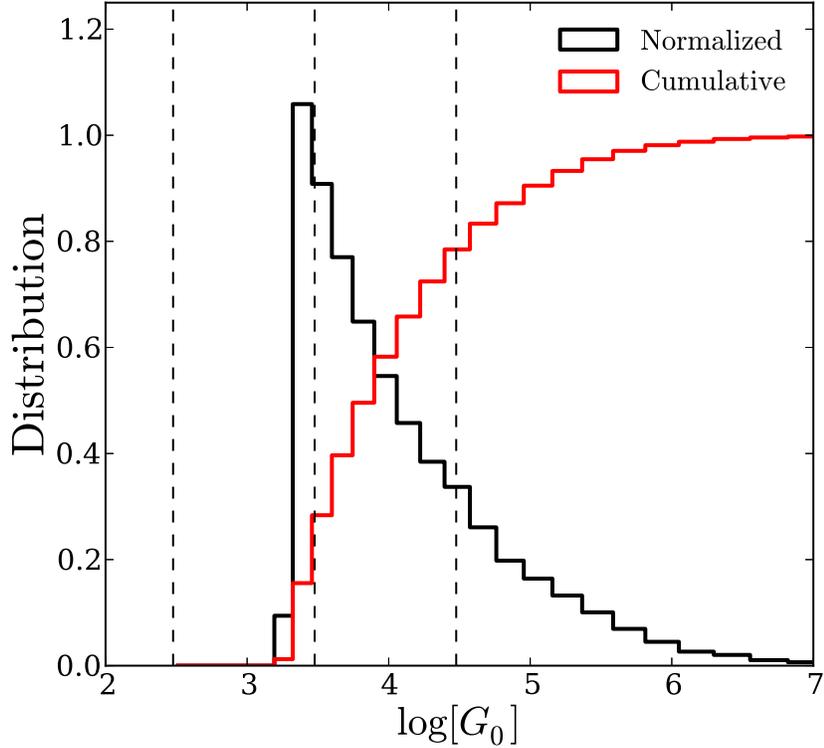}
\caption{Distribution of expected FUV fluxes based on the 28 disks in 
the ONC with measured masses and radii (denoted by blue data points in
Figure \ref{fig:isochrone}). Note that the scale for the $x$-axis 
uses common logarithms. For each disk with projected distance 
$d_p$ from the cluster center, the line-of-sight distance $s$ was
randomly sampled $N_s = 10^{5}$ times. For reference, the figure
includes benchmark values (marked by vertical dashed lines) for the
three FUV fluxes explored in this paper, i.e., $G_0$ = 300 (left),
3000 (middle), and 30,000 (right).  No disks are exposed to flux
levels $G_0 < 2000$ because of our estimate for $L_{\rm{FUV}}$, and
because all disks are assumed to be located within $r \leq 2.5$ pc 
of the cluster center.  The peak of the distribution occurs near 
$G_0 \approx 3000$, consistent with our previous assumptions.} 
\label{fig:G_dist}
\end{figure}

The results of this analysis are shown in Figure \ref{fig:G_dist},
where the random variable $s$ (from equation [\ref{s}]) has been
uniformly sampled $N_s = 10^{5}$ times for each of the 28 disks with
measured masses and radii (corresponding to the blue data points in
Figure \ref{fig:isochrone}).  Nearly $80\%$ of all disks in this
distribution have flux levels of $G_0 \leq 30,000$, indicated by the
rightmost vertical dashed line.  No disks in this sample have flux
levels below $G_0 \approx 2000$; this minimum flux occurs because all
are assumed to be located within the cluster radius $r_c = 2.5$ pc.
These results are consistent with the evolutionary tracks shown in
Figure \ref{fig:isochrone}, because the best agreement between the
model predictions occurs for higher FUV fluxes, those with 
$3000 \leq G_0 \leq 30000$. 

\section{Conclusion}

\subsection{Summary of Results}

In order to understand the effects of external stellar radiation on
disk evolution, this paper has incorporated photoevaporation models
\citep{adams} into time-dependent evolutionary models for the disks
(using an $\alpha$ prescription). We use this framework to derive
constraints on disk lifetimes, mass loss rates, radii, and masses. In
contrast to most previous work concerning photoevaporating disks, this
paper focuses on the effects of radiation fields from external stars
(but see also \citealt{armitage2000} and \citealt{clarke2007}).  We
also consider photoevaporation by X-rays from the host star
\citep{owen11b} in order to assess the relative importance of internal
versus external radiation sources for evaporation. An immediate
application (and test) of the model is provided by the Orion Nebula
Cluster, where the entire disk population is illuminated by the
Trapezium O stars, and where new data are available. Our main results
can be summarized as follows:

[1] Photoevaporation from external FUV sources severely reduces the
disk masses, and truncates the disk radii, over most of the expected
parameter space. For a given external radiation field, the most
important parameter in determining the disk lifetime is the viscosity
(given here by $\alpha$).  Disks with viscosity parameter in the range
$10^{-3} \lesssim \alpha \lesssim 10^{-2}$ diffuse outward on time
scales that are short enough so that disk material is continuously
transported to regions of low gravitational potential where it can
freely escape.  Even disks exposed to moderate FUV radiation ($G_0$ =
300) can be significantly depleted on short time scales when the
viscosity is sufficiently high ($\alpha \gtrsim 10^{-3}$).  Disks with
moderate viscosity (with $\alpha = 10^{-3}$) are effectively dispersed
in 1 -- 3 Myr; disks with higher viscosity ($\alpha = 10^{-2}$) are
dispersed within $\sim 0.5$ Myr (see Figures \ref{fig:mass} and
\ref{fig:G0_time}).  Disk sizes are quickly truncated to radii
$r_d<100$ AU over the full range of parameter space, and many disks 
are truncated even further, so that $r_d < 30 - 50$ AU (Figure
\ref{fig:edge_all}).

[2] When combined with existing X-ray photoevaporation models from the
host star \citep{owen11b}, the mass loss from external FUV radiation
dominates except for host stars with the highest X-ray luminosities
($L_X = 10^{31}$ \ergs).  Disk masses are usually depleted by external
FUV fields before any interesting effects from the X-rays can occur,
such as ``holes'' in the inner disk, i.e., regions where
photoevaporation clears out mass faster than it can be replenished
through viscous transport.  This kind of structure can only emerge
when the viscous time scale is comparable to the mass loss time scale
(for X-ray evaporation).  Since the viscous time scale starts out
relatively high, the disk must evolve for some time (typically $\sim
4-5$ Myr) before gaps can be formed by X-rays.

[3] Our evolutionary model (including only FUV radiation from external
stars, and excluding X-rays from the central star) is in good
agreement with observed masses and radii for disks in the Orion Nebula
Cluster (see Figure \ref{fig:isochrone}).  The best agreement with the
data generally occurs for relatively high radiation fields, those with
$G_0 = 3000 - 30,000$, and for disk viscosity parameters in the range
$10^{-3} \leq \alpha \leq 10^{-2}$. Figure \ref{fig:isochrone} also
shows that circumstellar disks follow well-defined tracks in the plane
$(r_d,M_d)$ of disk mass and radius; such plots can thus be used to
study disk evolution (analogous to the use of the H-R diagram for
stellar evolution).

\subsection{Discussion}

The main result of this paper is that disks can be readily destroyed
by strong external FUV radiation fields, sometimes within 1 Myr, often
within 2--3 Myr, and nearly always within 10 Myr. For fixed initial
disk masses and radii, this wide range in possible disk lifetimes is
due primarily to variations in disk viscosity ($\alpha$).  Since the
formation time scale of giant planets via core accretion is estimated
to lie in the range $\sim1-10$ Myr \citep{pollack96}, the potential
for planet formation in populated star-forming regions can often be
suppressed. This constraint can be evaded if the viscosity is
relatively low (with $\alpha \lesssim 10^{-3}$).  At first glance,
this result is puzzling, because most stars are thought to form in
clusters rather than in isolation \citep{lada2003} -- and yet both
single and multiple planet systems appear to be common. Furthermore,
our own Solar System had sufficient material from which to form
planets, even though it was probably born in a relatively populated
cluster \citep [and references therein]{adams_araa}.

However, it is important to remember that only a subset of young stars
experience the most extreme environments considered in this paper.
The regions of parameter space where planet formation is most
threatened are those with FUV fluxes $G_0 \sim 30,000$, which
generally corresponds to disks located within $ \sim 0.1$ pc of the OB
star(s) generating the FUV fields (the exact number will depend on the
luminosity of the FUV sources, which varies from cluster to cluster).
At such close distances, planet formation is likely to be compromised
unless the disk viscosity is extremely low. Most disks, however, are
located at distances $r \sim 1$ pc, so they experience lower flux
levels (even in populated clusters). The density profile of a star
cluster can be approximated as $\rho(r) \sim r^{-2}$, so that the
enclosed mass $M(r) \propto r$.  As a result, most of the stars (and
their disks) are located in the outer regions of the cluster, and are
therefore exposed to fluxes near the lower end of the range considered
in this paper.  If we take $G_0 = 300$ as a typical flux value for
many disks, then planet formation is probably not significantly
threatened, as long as there are enough disks with viscosity
parameters smaller than $\alpha = 10^{-2}$.

Another way for disk lifetimes to be extended long enough to allow
planet formation is by considering the dynamical interactions of
star-disk systems within clusters.  Over the course of $\sim1$ Myr,
stars can change their positions relative to the massive star(s) that
are responsible for most of the radiation.  If a star's orbit around
the photon source is highly eccentric\footnote{Note that describing
  orbits as ``eccentric'' here is somewhat misleading, or at least
  requires further specification, because the gravitational potential
  of the cluster is not Keplerian.}, it will spend most of its time in
the outer cluster regions with low FUV flux, and experience only brief
periods of higher flux.  This could potentially allow individual disks
to retain their mass for somewhat longer, perhaps long enough to form
planets.  Considering these types of dynamics would be equivalent to
introducing a time-dependent $G_0$ into our model.  Note however that
if the field strength never falls below $G_0 \sim 300$ over an orbit,
then the results of this paper imply that dynamical interactions can
probably only extend disk lifetimes by a factor of two or so.

Although the implications for planet formation remain somewhat
ambiguous, in that they depend strongly on the disk viscosity, the
results of this paper show that planets with large semimajor axes
(those with $a \approx 100$ AU) should be rarely produced in heavily
populated star clusters.  After 1 Myr have elapsed, disks are either
completely dispersed, or are truncated to radii $r_d < 100$ AU (see
Figure \ref{fig:edge_all}).

Future research can be taken in several directions.  The
time-dependent disk model developed in this paper can be used to study
a wider variety of systems, with parameters outside of the range
explored herein.  In particular, we have only considered systems with
host stars $M_* = 1 \msun$.  Considering stars with larger masses is
especially important, because massive stars produce deeper
gravitational potential wells, which could allow their disks to resist
being photoevaporated.  As mentioned previously, one goal of this
paper is to identify the relative importance of internally versus
externally generated radiation fields.  This paper has made progress
in addressing this issue, but much work still remains.  It is
important to emphasize that the most important factor in this issue is
the environment in which the disk resides -- many disks living in
relative isolation will never be exposed to the damaging
photoevaporating fields discussed in this paper, and hence this
discussion is restricted to objects in more populated clusters. We
have shown that for disks around solar-mass stars, the effects of
X-ray illumination from the host star are usually insignificant
compared to the effects of external FUV illumination.  However, we
have not considered FUV radiation from the host star itself.
\cite{gorti2} find that for a viscosity of $\alpha = 10^{-2}$, disks
are dispersed within $\sim 4$ Myr, and for $\alpha = 10^{-3}$ disk
lifetimes can exceed 10 Myr.  For comparison, our model predicts that
disks with $\alpha = 10^{-2}$ are effectively dispersed in less than
0.5 Myr and disks with $\alpha = 10^{-3}$ within 1-3 Myr (see Figure
\ref{fig:G0_time}).  Comparing these time scales for disk dispersal,
FUV fields from the host star are not expected to drastically alter
the results of this paper, but they could provide corrections of,
perhaps, factors of order unity, and should be included in future
calculations.

\acknowledgments

This paper benefited from discussions with many colleagues, especially
Jaehan Bae, Konstantin Batygin, and James Owen.  We also thank an
anonymous referee for useful comments. KRA and NC were supported in
part by NASA Origins Grant NNX08AH94G; FCA was supported in part by
NASA Origins grant NNX11AK87G.

\end{document}